\documentclass{aastex631} 

\submitjournal{ApJ}

\shorttitle{Magnetic helicity fluxes from triple correlators}
\shortauthors{Gopalakrishnan and Subramanian}

\let\oldtablenum\tablenum 
\let\tablenum\relax
\usepackage{siunitx}
\let\tablenum\oldtablenum

\usepackage{amsmath}
\usepackage{mathtools} 
\allowdisplaybreaks 

\usepackage[inline]{enumitem} 

\usepackage[normalem]{ulem}
\usepackage{cancel} 

\renewcommand{\dh}{\partial}
\renewcommand{\d}{\mathrm{d}}
\newcommand{\mean}{\overline}
\renewcommand{\vec}[1]{\boldsymbol{#1}}
\newcommand{\meanvec}[1]{\boldsymbol{\mean{#1}}}
\newcommand{\unitvec}[1]{\boldsymbol{\hat{#1}}}
\newcommand{\op}{\mathrm}
\newcommand{\eps}{\epsilon}
\newcommand{\emf}{\mathcal{E}}
\newcommand{\defn}{\equiv}
\newcommand{\bigO}{\mathcal{O}}

\newcommand{\abs}[1]{\left|#1\right|}
\newcommand{\continuedTerm}{\phantom{+++}} 
\newcommand{\meanBr}[1]{\left<#1\right>}
\newcommand{\exchangeArrow}{\leftrightarrow}
\newcommand{\dive}{\nabla\cdot}
\newcommand{\taghere}{\stepcounter{equation}\tag{\theequation}} 
\newcommand{\curl}[1]{\nabla\times#1}
\newcommand{\cross}{\times}
\newcommand{\Rm}{\ensuremath{\operatorname{\mathrm{Rm}}}} 

\newcommand{\Eq}[1]{Eq.~(\ref{#1})}
\newcommand{\Eqs}[2]{Eqs.~(\ref{#1}) and~(\ref{#2})}

\newcommand{\Eqsn}{Eqs.\@}

\begin{document}

\title{Magnetic helicity fluxes from triple correlators}

\correspondingauthor{Kishore Gopalakrishnan}
\email{kishoreg@iucaa.in}

\author[0000-0003-2620-790X]{Kishore Gopalakrishnan}
\affiliation{IUCAA, Post Bag 4, Ganeshkhind, Pune 411007, India}

\author[0000-0002-4210-3513]{Kandaswamy Subramanian}
\affiliation{IUCAA, Post Bag 4, Ganeshkhind, Pune 411007, India}

\begin{abstract}
		Fluxes of the magnetic helicity density play an important role in large-scale turbulent dynamos, 
		allowing the growth of large-scale magnetic fields while overcoming catastrophic quenching.
		We show here, analytically, how several important types of magnetic helicity fluxes
		can arise from terms involving triple correlators of fluctuating fields in the helicity density evolution equation.
		For this, we assume incompressibility and weak inhomogeneity, and use a quasinormal closure approximation:
		fourth-order correlators are replaced by products of second-order ones, and 
		the effect of the fourth-order cumulants on the evolution of the third moments is modelled by a strong damping term. 
		First, we show how a diffusive helicity flux, till now only measured in simulations, arises from the triple correlation term.
		This is accompanied by what we refer to as a `random advective flux', which predominantly transports magnetic helicity along the gradients
		of the random fields.
		We also find that a new helicity flux contribution, in some aspects similar to that first proposed by Vishniac, can arise from the triple correlator.
		This contribution depends on the gradients of the random magnetic and kinetic energies along the large-scale vorticity,
		and thus arises in any rotating, stratified system, even if the turbulence is predominantly nonhelical.
		It can source a large-scale
		dynamo by itself while spatially transporting magnetic helicity within the system.
\end{abstract}


\section{Introduction}

Dynamo theory studies the spontaneous amplification and maintenance of  magnetic fields by electromagnetic induction due to the motion of a conducting fluid.
Astrophysical magnetic fields in stars and galaxies are observed to be ordered on scales much larger than the integral scale of the fluid turbulence that is partially responsible for their maintenance \citep{KanduPhysReview,BeChElBl19}.
Large-scale (or mean-field) turbulent dynamo theory attempts to understand the mechanisms behind the generation and sustenance of such fields, which typically involves some form of breaking of mirror symmetry.
In the standard picture, differential rotation generates toroidal
magnetic fields from poloidal ones, while helical motions re-generate the poloidal field from the toroidal field, by what is referred to as the $\alpha$-effect
\citep{Mof78,KR80,KanduPhysReview,rincon_2019,ss21,Tobias21}.

Dynamos saturate when Lorentz forces due to the generated magnetic fields back-react on the velocity fields driving the dynamo. 
In mean-field helical dynamos, this saturation is strongly constrained by the 
near conservation of 
magnetic helicity (a measure of links, twists and writhing of the field).
In such dynamos, the generation of a large-scale magnetic field 
and the associated
magnetic helicity is accompanied by
concurrent
transfer of an equal amount of magnetic helicity of the opposite sign to the small scales.
This transfer 
is accomplished by the turbulent electromotive force, the cross correlation
between the fluctuating velocity and magnetic fields (see below).
Such a `bihelical' field has indeed been found in observations and simulations \citep{BB03,Singh_2018}.

Although equal and oppositely signed magnetic helicities accumulate at large and small scales as the mean field grows, 
the Lorentz force is dominated by the smaller scales. 
In simple closures, one can show that this leads to a small-scale current helicity that opposes the effect of helical motions by decreasing the kinetic $\alpha$-effect.
A drastic suppression of the kinematic $\alpha$-effect, which would also suppress mean-field growth,
is referred to as `catastrophic' quenching \citep{CV91,VaCa92}. This was attributed to helicity conservation by \citet{GD,BY95}.
Thus, 
one requires some mechanism for shedding the accumulated small-scale helicity.
Resistive dissipation is too slow.
However, 
fluxes of 
unwanted 
small scale helicity out of the dynamo active region
can 
allow the mean field dynamo to grow on timescales much smaller than the Ohmic timescale 
\citep{BF00,KanduPhysReview}.

A number of such fluxes have been derived 
by simplifying various terms in the evolution equation for the small-scale magnetic 
helicity \citep{kleeorin2000MFD,vishniac2001magnetic,Kandu2004helicityFluxes,Kandu2006helicityFlux,VS14,ss21,Kleeorin22}.
They are used in dynamical quenching models, where the mean field dynamo equation is solved together with
the evolution equation for the current helicity, to show how quenching of the large-scale dynamo can be alleviated 
\citep{kleeorin2000MFD,SSS07,CSS13a,CSS15}.
The helicity fluxes considered so far mostly
require either large-scale outflows, or a large-scale magnetic field, to operate.
Astrophysical systems need not host such outflows.
Further, a flux which depends on the mean magnetic field is not expected to be important in the initial stages of dynamo action.
There is thus a need to explore if other fluxes of magnetic helicity are possible which can alleviate catastrophic quenching or even lead to new generative effects.

One helicity flux term which has so far not been studied as systematically as other terms involves a triple correlation of small-scale fields. 
Its simplification requires a closure approximation, and its study is the main focus of the current paper. 
We show that simplification of this triple correlation term leads to several important helicity fluxes. 
First, it leads to a diffusive flux of the magnetic helicity, 
so far only postulated (as a flux of current helicity) based on heuristic arguments \citep[e.g.][]{Covas1998MFD, kleeorin2002MFD}
and measured in direct numerical simulations \citep{MitraDiffusiveSim}.
We find that this flux comes together with an advective flux which depends on gradients of the random fields.
We also show that a helicity flux contribution, similar to
that first proposed by \citet{newVishniacFlux}, and depending only on small-scale field gradients and large-scale vorticity, 
can arise from this triple correlator term.
Indeed this latter flux could potentially drive a large-scale dynamo by itself, even without the presence of kinetic helicity.

In section \ref{section: helicity evolution}, we set up the mean-field formalism and describe the evolution equation for the small-scale magnetic helicity.
In section \ref{section: eval triple correlator flux}, we simplify the triple correlator using a closure approximation, 
presenting a partially simplified expression (which nevertheless already displays all the relevant effects) in section \ref{section: triple flux local terms}.
In section \ref{section: nonlocal terms simplification}, we apply a further approximation to simplify some nonlocal terms involving the pressure and the scalar potential. 
The detailed expressions for many of the correlators involved are left to the appendices.
In section \ref{section: physical implications}, we discuss some physical implications of the fluxes we have obtained.
Finally, section \ref{section: conclusions} presents our
conclusions.

\section{Magnetic helicity: evolution and fluxes}
\label{section: helicity evolution}
Let us consider an incompressible fluid, and use both overbars and angle brackets to denote an average that follows Reynolds' rules.
The magnetic field, $\vec{B}$, evolves according to the induction equation,
\begin{equation}
	\frac{\dh\vec{B}}{\dh t} = \curl\left(\vec{V}\cross\vec{B} - \eta \curl\vec{B}\right)
	\,,
	\label{eq: induction}
\end{equation}
where $\vec{V}$ is the velocity and $\eta$ is the resistivity.
We now average \Eq{eq: induction} after splitting all quantities into their mean and fluctuating parts (e.g.\@ $\vec{B} = \meanvec{B} + \vec{b}$ with $\meanBr{\vec{b}} = 0$).
Here, fluctuating quantities are denoted by lowercase letters.
The evolution equation for the mean magnetic field is then
\begin{equation}
	\frac{\dh\meanvec{B}}{\dh t} = \curl \left( \meanvec{V}\cross \meanvec{B} + \vec{\emf} - \eta \curl \meanvec{B} \right) 
	\label{eq: mean field induction}
\end{equation}
where $\vec{\emf} \equiv \meanBr{\vec{v}\times\vec{b}}$ is called the \emph{turbulent electromotive force (EMF)}. 

Expressing $\vec{\emf}$ in terms of the mean fields themselves
is a closure problem. 
Assuming a sufficient scale separation between the mean and fluctuating fields, $\vec{\emf}$ is expanded in terms of the mean magnetic field and its derivative.
For statistically homogeneous and isotropic small-scale fields, 
one then gets
\begin{equation}
	\vec{\emf}
	=
	\alpha \meanvec{B}
	- \eta_t \curl\meanvec{B}
	\,.
	\label{eq: emf alpha etat}
\end{equation}
One can use
the quasilinear approximation to show that $\alpha$ is related to the kinetic helicity of the fluctuating velocity field
when Lorentz forces are 
negligible, 
while $\eta_t$, the turbulent resistivity, depends on its energy density \citep{Mof78}.
A nonzero $\alpha$ allows a mean magnetic field to grow and be sustained even without a mean velocity field.

Subtracting \Eq{eq: mean field induction} from \Eq{eq: induction},  we obtain 
the evolution equation for the fluctuating magnetic field:
\begin{equation}
	\frac{\dh b_i}{\dh t} = - \eps_{ijk}\dh_je_k
	\,,
	\label{eq: induction random in terms of e}
\end{equation}
where
\begin{equation}
	e_k \defn 
	- \eps_{klm} \mean{V}_lb_m 
	- \eps_{klm}v_l\mean{B}_m 
	- \eps_{klm}v_lb_m 
	+ \emf_k 
	+ \eta\eps_{klm}\dh_lb_m
	\,,
	\label{eq: e definition}
\end{equation}
so that the fluctuating electric field is given by $\vec{e}/c$.
The fluctuating vector potential (which we denote by $\vec{a}$) then evolves according to
\begin{equation}
	\frac{\dh a_k}{\dh t} =  - e_k + \dh_k\varphi
	\,,
	\label{eq: random vector potential time evolution}
\end{equation}
where $\varphi$ is a scalar potential (our definition differs from the usual one by a negative sign).
The evolution of the random velocity field $\vec{v}$ is given by
\begin{equation}
	\begin{split}
		\frac{\dh v_i}{\dh t} ={}& - \mean{V}_j\dh_jv_i - v_j\dh_j\mean{V}_i - v_j\dh_jv_i + \meanBr{ v_j\dh_jv_i } 
		-\frac{\dh_i p}{\rho} + \frac{\mean{B}_j\dh_jb_i}{4\pi\rho} + \frac{b_j\dh_j\mean{B}_i}{4\pi\rho} + \frac{b_j\dh_jb_i}{4\pi\rho} - \frac{\meanBr{ b_j\dh_jb_i }}{4\pi\rho} + \nu\dh_j\dh_jv_i
		\,.
		\end{split}\label{eq: NS random}
\end{equation}
Here, $\rho$ is the density (taken as constant assuming an incompressible flow), $\nu$ is the viscosity, and $p$ is the pressure determined using $\nabla  \cdot \vec{v} =0$ in \Eq{eq: NS random}.

We
define the magnetic helicity density as 
$h^b \equiv \meanBr{\vec{a}\cdot\vec{b} }$,
and work in the Coulomb gauge, with $\dive\vec{a}=0$.
This 
implies that $\nabla^2\varphi = \dh_k e_k $.
Using \Eqs{eq: induction random in terms of e}{eq: random vector potential time evolution}, 
we can write the time derivative of the magnetic helicity density as
\begin{equation}
	\frac{\dh h^b}{\dh t} = -2 \meanBr{e_ib_i } + \meanBr{b_i\dh_i\varphi } - \dh_j\meanBr{\eps_{jki}e_ka_i }
	\,.
	\label{eq: hb evolution in terms of e}
\end{equation}
Substituting \Eq{eq: e definition} into the above, we write
\begin{align}
	\begin{split}
		\frac{\dh h^b}{\dh t} 
		={}&
		- 2 \, \vec{\emf} \cdot\meanvec{B} 
		- 2 \, \eta \meanBr{ \vec{b} \cdot \left( \curl\vec{b} \right) }
		+ \meanBr{b_i\dh_i\varphi } 
		+ \overbrace{ \dh_j{\left( \meanBr{\eps_{jki} \eps_{klm} b_m  a_i } \mean{V}_l \right)} }^\text{advective} {}
		\\& + \underbrace{\dh_j{\left( \meanBr{\eps_{jki} \eps_{klm} v_l a_i } \mean{B}_m \right)} }_\text{A+VC} {}
		+ \underbrace{ \dh_j\meanBr{\eps_{jki} \eps_{klm}v_lb_m a_i } }_\text{triple-correlator} {}
		-\dh_j \left(\eta\meanBr{\eps_{jki}a_i \eps_{klm}\dh_lb_m }\right)
		.
	\end{split} \label{eq: helicity evolution full}%
\end{align}

The first term above transfers magnetic helicity between the mean and fluctuating fields, while the second describes
resistive dissipation of magnetic helicity density.
The term marked `advective' 
corresponds to advection of the magnetic helicity by the mean velocity field 
in the isotropic limit \citep{Kandu2006helicityFlux}.
The term marked `A+VC' can be split into two parts, which are known as the `antisymmetric' \citep{kleeorin2000MFD} and `Vishniac-Cho' fluxes  \citep{vishniac2001magnetic,Kandu2006helicityFlux}.
These are dependent on the mean magnetic field, and are thus unimportant when the mean magnetic field is weak (i.e.\@ in the initial stages of dynamo action).
The term marked 'triple correlator' requires a closure hypothesis to evaluate, and has not yet
been studied.
The last term in \Eq{eq: helicity evolution full} is a resistive flux, 
expected to be negligible 
at high magnetic Reynolds numbers, $\Rm$.
Finally, 
$\meanBr{b_i\varphi}$
is a non-local term which gives contributions similar to all the above fluxes.

Suppose the turbulent quantities are statistically steady, and $\dh h^b/\dh t = 0$, in a time averaged sense. Then \Eq{eq: helicity evolution full} implies
\begin{equation}
		\vec{\emf} \cdot\meanvec{B} 
		= 
		- \eta \meanBr{ \vec{b} \cdot \left( \curl\vec{b} \right) }
		- \frac{1}{2} \, \dive{\left( \text{fluxes} \right)}
		\,.
		\label{eq: steady emf}
\end{equation}
In the absence of magnetic helicity fluxes, $\vec{\emf}\cdot\meanvec{B}$ is dominated by 
the resistive term, 
and becomes negligible at high $\Rm$.
The mean field dynamo
then becomes catastrophically quenched in the absence of helicity fluxes.
Thus, 
helicity fluxes are crucial for the growth of the mean magnetic field over timescales shorter than the resistive one.

In this paper, we will focus on the hitherto unevaluated contribution to \Eq{eq: helicity evolution full} marked as `triple correlator', which arises from the $-\vec{v}\times\vec{b}$ part of $\vec{e}$, and also similar terms which arise from $b_i\varphi$.

\section{Evaluation of the flux due to the triple correlator}
\label{section: eval triple correlator flux}

\subsection{Simplification of the local terms}
\label{section: triple flux local terms}
The term in \Eq{eq: e definition} for the small-scale electric field $\vec{e}$ that we are interested in is $- \vec{v}\times\vec{b}$.
Substituting this in \Eq{eq: hb evolution in terms of e}, the triple correlator contribution to helicity density evolution is given by
\begin{equation}
	\left(\frac{\dh h^b}{\dh t}\right)_{\text{triple}} =
	\underbrace{\dh_j\meanBr{v_ib_ja_i }}_{I^\text{triple}_1} - \underbrace{\dh_j\meanBr{v_jb_ia_i }}_{I^\text{triple}_2}
	.
	\label{eq: helicity evolution}
\end{equation}

The two terms $I^\text{triple}_1$ and $I^\text{triple}_2$ contain triple correlators involving the random velocity, magnetic, and vector potential fields.
These 
are evaluated in appendix \ref{appendix: triple correlator evaluation} by the following 
steps:
\begin{enumerate}
	\item
	We write the evolution equations for the triple correlators by taking their time derivatives.
	These contain fourth-order correlators along with triple correlators involving the pressure or the scalar potential.
	
	\item
	We assume that the fourth-order correlators 
	can be written in terms of second-order correlators (which would be the case if the turbulent fields were all Gaussian) plus a damping term with some characteristic timescale $\tau$, that models the irreducible fourth order cumulant.
	The 4th-order correlator also involves the product of third and first order correlators, but this vanishes
	as fluctuations have zero mean.
	This is similar to the Eddy Damped Quasi-Normal (EDQN) closure \citep{Lesieur08}. 
	We further assume that this damping timescale is much smaller than the slower evolution timescales of various averaged quantities we are interested in, so that we can drop the time derivatives in the evolution equations for the triple correlators.
	\item
	Contributions to the triple correlation flux involving the EMF and the mean magnetic field are neglected.
	These terms  
	are negligible in the initial stages of dynamo action, when an appreciable large-scale field has not 
    yet 
    developed.
	Terms involving 
	$\eta$ or 
	$\nu$ are neglected since the magnetic and fluid Reynolds numbers are high in most astrophysical systems.
	\item
	At this stage, one has an expression for the triple correlators in terms of the double correlators of various fields (except for non-local terms that involve 
	$p$ or $\varphi$).
	To simplify these double correlators, we use the expressions in appendix \ref{appendix: RobertsSoward expressions}, 
	which are
	from \citet{robertsSoward}.
	These correspond to assuming incompressibility and weak inhomogeneity.
	The resulting 
	double-correlators are listed in appendix \ref{appendix: double correlators real-space expressions}.
	They match those 
	derived using 
	an alternative method proposed by \citet{ZeldovichFieldSplit}.
\end{enumerate}

Substituting the final expressions for $I^\text{triple}_1$ from \Eq{eq: I1 final} and $I^\text{triple}_2$ from \Eq{eq: I2 final} into \Eq{eq: helicity evolution}, we obtain \Eq{eq: dhbdt unsimplified pressure phi}.
There are diffusive and advective fluxes, along with terms similar to a flux previously proposed by \citet{newVishniacFlux, Vishniac2015}.
The physical interpretation of these fluxes will be discussed in section \ref{section: physical implications}.
There are also some unsimplified terms involving the pressure or the scalar potential.
These terms may modify the coefficients and properties of the fluxes, and so we now deal with them.

\subsection{Simplification of the nonlocal terms}
\label{section: nonlocal terms simplification}

The evaluation of these terms is more involved, as they depend on
an integral operator (the inverse of the Laplacian) acting on the correlators between fields at spatially separated points.
We approximate these terms by replacing nonlocal integrals by appropriate powers of an eddy correlation length, which we denote by $\lambda$ \citep{vishniac2001magnetic}.
Also, for simplicity, we take the same constant $\lambda$ for all correlators.
For example, to estimate a term of the form $- \meanBr{ \dh_a\left[ \nabla^{-2}(f(\vec{x}) )\right] g(\vec{x})}$,
one could write it as
\begin{align}
	\begin{split}
	  &	\int\d\vec{y} \, \frac{ y_a - x_a }{4\pi \abs{\vec{x} - \vec{y}}^3} \, \meanBr{f(\vec{y}) \, g(\vec{x})}
		={}
		\int\d\vec{r} \, \frac{ r_a }{4\pi \abs{\vec{r} }^3} \, \meanBr{f(\vec{x}+\vec{r}) \, g(\vec{x})}
	\end{split}
	\\
	\begin{split}
		\approx{}&
		\int\d\vec{r} \, \frac{ r_a }{4\pi \abs{\vec{r} }^3} \, \meanBr{g(\vec{x}) \left( f(\vec{x}) + r_b \frac{\dh f(\vec{x})}{\dh r_b} + \frac{r_a r_b}{2}\, \frac{\dh^2 f}{\dh r_a \dh r_b} + \dots \right)}
		\approx{}
		\frac{\lambda \delta_{ab} }{3} \meanBr{g(\vec{x}) \frac{\dh f(\vec{x})}{\dh x_b}}
		+ \bigO(\lambda^3)
		\,.
	\end{split}
\end{align}

This approximation is applied in Appendix \ref{appendix: dpba correlator} 
and Appendix \ref{appendix: potential correlators} 
to evaluate the nonlocal terms in \Eq{eq: dhbdt unsimplified pressure phi} involving the pressure and scalar potential 
respectively.
The Vishniac-Cho (VC) approximation is also used
to evaluate the nonlocal
term $\meanBr{b_i\varphi}$ in Appendix \ref{section: appendix bphi correlator}. For the latter, we keep only contributions to $\vec{e}$ from 
$-\overline{\vec{V}} \times \vec{b}$ and $\vec{v} \times \vec{b}$, neglecting again the terms depending on the mean magnetic field and
resistivity. 
The calculation involves evaluating several nonlocal triple correlators, and the 
result is given by \Eq{eq: bphi correlator contribution}.
Some of the terms here are similar to
those resulting from the advective flux, which itself is calculated in Appendix \ref{appendix: advective flux}.

We use $-\dh_j(F^T_j)$ to denote the total contributions to 
$\partial h^b/\partial t$ (\Eq{eq: helicity evolution full})
from triple correlations, 
the $\meanBr{b_i\phi}$ and advection terms. This expression is quite cumbersome, and
is given in Appendix \ref{appendix: dhbdt nonlocal with hv and lam^2} 
(except for terms that are higher-order, $\bigO(\tau^2)$, corrections to the advective flux). 
To make this expression more tractable, we replace
$\lambda^2 \meanBr{\omega^2} = \meanBr{v^2}$, $\lambda^2 \meanBr{j^2} = \meanBr{b^2}$, 
$\lambda^2 \meanBr{b^2} = \meanBr{a^2}$,
$ \lambda^2 \int\d k \, 8\pi k^4 N(k,\vec{R}) = h^c$, and $ \lambda^2 \int\d k \, 8\pi k^4 F(k,\vec{R}) =h^v$.
This assumes that there is a dominant scale $\lambda$ in all the spectra.
If we are interested only in regimes where the kinetic $\alpha$-effect contributions 
in the helicity flux are minimal, we can drop terms involving $h^v$.
Then, 
we get a total helicity flux $\vec{F}^T$, where 
\begin{align}
	\begin{split}
		-\dh_j(F^T_j)
		=
		\dh_j\Bigg[ &
			\overbrace{
			\frac{7\tau}{27} \meanBr{v^2} \dh_j h^b
			+ \frac{7\tau}{27} \frac{\meanBr{b^2}}{4\pi\rho} \dh_j h^b
			}^{\text{diffusion}} {}
			- \overbrace{
			\frac{\tau}{18} h^b \dh_j \meanBr{v^2 }
			- \frac{7 \tau}{27 } \frac{1}{4\pi\rho} h^b \dh_j \meanBr{b^2 }
			}^{\text{random advection}}
			\\&
			- \overbrace{
			\frac{13 \tau^2 }{135} \frac{1}{4\pi\rho} \eps_{jkl} \mean{V}_l \dh_k \left( h^b h^c \right)
			- \frac{\tau^2 }{18} \frac{1}{4\pi\rho} \eps_{jkl} \mean{V}_l h^c \dh_k h^b 
			}^{\text{random advection}} {}
			+ \overbrace{
			\frac{7 \tau^2 }{45} \frac{1}{4\pi\rho} \epsilon_{jkl} \mean{V}_l \meanBr{b^2} \dh_k\meanBr{b^2 } 
			}^{\text{NV}}
			\\& - \underbrace{
			\frac{203 \tau^2 }{5400} \epsilon_{jkl} \mean{V}_l \meanBr{b^2} \dh_k\meanBr{v^2 }
			+ \frac{403 \tau^2 }{8100} \epsilon_{jkl} \mean{V}_l \meanBr{ v^2 } \dh_k \meanBr{b^2} 
			- \frac{\lambda^2}{6}  \eps_{jkl} \mean{V}_l  \dh_k \meanBr{b^2 }
			}_{\text{NV}} {}
			- \mean{V}_j h^b
		\Bigg]
		.
	\end{split} \label{eq: dhbdt with nonlocal simplified}%
\end{align}

\section{Implications of the obtained fluxes}
\label{section: physical implications}
We write the total helicity flux $\vec{F}^T$ in \Eq{eq: dhbdt with nonlocal simplified} as a sum of a diffusive flux $\vec{F}^D$, random advective flux $\vec{F}^{RA}$, advective flux $\vec{F}^A$ 
and NV flux $\vec{F}^{NV}$. 
The terms which involve $h^bh^c$ also contribute to these fluxes. 
We note that the terms which involve the density $\rho$ 
arise from the small-scale Lorentz force.
We now discuss these fluxes in turn.

\subsection{The diffusive and advective fluxes}
Diffusive fluxes of the magnetic helicity have been used in models of the mean-field dynamo which include a dynamical equation for the current helicity of the small-scale fields \citep{Covas1998MFD, kleeorin2002MFD}.
Such a flux has been measured in direct numerical simulations \citep{MitraDiffusiveSim}.
In some situations, it has been found that the inclusion of such a diffusive term in a mean-field model helps overcome quenching of the dynamo \citep{brandenburg2009}.
This flux is derived here for the first time, and results from the triple correlation contribution
to the helicity fluxes.
In many physical contexts, random motions can lead to not only diffusion, but also advection due to their gradients.
This is also true here, where we see from \Eq{eq: dhbdt with nonlocal simplified} an advective flux which depends on gradients
of $\meanBr{v^2}$ and $\meanBr{b^2}$, along with a term involving $\vec{V}$ and $h^c$.
The explicit form of these two fluxes is 
\begin{equation}
     \vec{F}^D = 
		- \left(\frac{7\tau}{27} \meanBr{v^2}
		+ \frac{7\tau}{27} \frac{\meanBr{b^2}}{4\pi\rho}\right) \nabla  h^b 
		\,, \quad
	\vec{F}^{RA} =
		h^b \left(
			\frac{\tau}{18} \nabla \meanBr{v^2 }
			+ \frac{7\tau}{27} \frac{1}{4\pi\rho} \nabla  \meanBr{b^2 } 
			+ \frac{\tau^2}{18} \frac{1}{4\pi\rho} \left[
				\meanvec{V} \times \nabla h^c 
				- \frac{41}{15}h^c\curl \meanvec{V}
			\right]
		\right)
		.
	\label{eq: diffusive flux}
\end{equation}
The random velocity contributes a diffusion coefficient of order $\eta_t$ (where $\eta_t = \tau\meanBr{v^2 }/3$) in $\vec{F}^D$; at equipartition,
the random magnetic field doubles this value. Near equipartition, we also find that the magnetic term dominates the random advection. 
Comparing $\vec{F}^{RA}$ with advective flux ($\vec{F}^A = \meanvec{V} h^b$, the last term of \Eq{eq: dhbdt with nonlocal simplified}), we note that even when $\meanvec{V}=0$, the helicity is advected by an effective velocity which is related to the gradients of the fluctuating 
fields.

\subsection{The New Vishniac Flux}

The terms marked `NV' (standing for `New Vishniac' Flux) are, in some aspects, similar to a flux previously proposed by \citet{newVishniacFlux, Vishniac2015}.\footnote{
The general form of this flux has been discussed by Vishniac in several talks (and in
private communications) but not yet published, apart from the abstracts cited 
\citep{newVishniacFlux,Vishniac2015}. So, unfortunately, a detailed comparison is not possible.
} 
These can be rewritten as
$- \dive\vec{F}^{NV}$, where
\begin{align}
 \begin{split}
   \vec{F}^{NV}
   ={}&
   \left( \curl \meanvec{V} \right)\bigg[ 
		C_1 \frac{\tau^2}{8\pi\rho} (\meanBr{b^2 })^2
		+ C_2 \tau^2 \meanBr{ v^2 } \meanBr{b^2} 
		+C_4 \lambda^2 \meanBr{b^2}
		\bigg]
		+\tau^2 \left( C_3 - C_2 \right) \bigg( \meanvec{V} \times \meanBr{v^2} \nabla \meanBr{b^2 } \bigg)
		\,,
 \end{split}   
    \label{eq: NVF flux rewrite}
\end{align}
and the constants are given by $(C_1, C_2, C_3, C_4 ) =(
{7 }/{45} ,- {203 }/{5400} ,{403 }/{8100} ,-{1}/{6})$.
We also get
\begin{align}
	\begin{split}
	-  \dive \vec{F}^{NV} 
		={}&
		- \left( \curl \meanvec{V} \right) \cdot \bigg[ 
		C_1 \frac{1}{4\pi\rho} \meanBr{b^2} \tau^2 \nabla \meanBr{b^2 } 
		+ C_2 \meanBr{b^2} \tau^2 \nabla\meanBr{v^2 }
		+ C_3 \meanBr{ v^2 } \tau^2 \nabla \meanBr{b^2} 
		+C_4 \lambda^2 \nabla  \meanBr{b^2}
		\bigg]
		\\& - \tau^2 \left( C_3 - C_2 \right) \meanvec{V} \cdot \bigg( \nabla \meanBr{b^2} \cross \nabla \meanBr{v^2 } \bigg)
		.
	\end{split} \label{eq: NVF flux rewrite div}
\end{align}

The expression in \Eq{eq: NVF flux rewrite}
is our final result for 
the NV flux,
and \Eq{eq: NVF flux rewrite div} 
is its contribution to the time evolution of the magnetic helicity density.
The NV flux $\vec{F}^{NV}$ has even parity, while $\dive\vec{F}^{NV}$ is explicity parity odd (like $h^b$).

We note in passing that \citet[eq.~18]{Kleeorin22} have calculated helicity fluxes
\emph{assuming} all triple correlations combine to contribute a purely diffusive flux governed by $\eta_t$. Thus, they do 
not get any of the other diffusive,
RA or NV terms in \Eq{eq: NVF flux rewrite div}, that we obtain from triple correlator 
contributions. They do get an NV flux term similar to the $C_4$ term that arises from double correlators in the
$\meanBr{b_i\phi}$ and advective terms of \Eq{eq: helicity evolution full}. 
They retain the effect of the mean magnetic field, and all
their fluxes (except the $C_4$ term) depend on this being strong.
Similarly, \citet[eq.~51]{pipin08} has also obtained a term similar to the $C_4$ term as a flux of the current helicity.

The second line in \Eq{eq: NVF flux rewrite div} can be neglected if we assume all the turbulent correlators 
vary predominantly in one direction.
This may be the case in a disk galaxy or in an accretion disk, where we expect 
a predominant stratification perpendicular to the disk; and in stars where we expect it to be radial.
The first line in \Eq{eq: NVF flux rewrite div} is non-zero in the presence of mean vorticity and stratification, and will lead to magnetic
helicity generation even if the initial random velocity and magnetic fields are non-helical.
This corresponds to a current helicity $h^c=h^b/\lambda^2$, which in simple closures leads to a magnetic 
alpha-effect $\alpha_m = (\tau/3)(h^c/4\pi\rho)$ \citep{PFL76,GD,BF02b,radler2003,KanduPhysReview}. 
Such terms could be particularly interesting
in magnetically dominated turbulence.
Rotation and stratification can also lead to kinetic helicity generation, whose
source has a similar mathematical form and in turn results in a kinetic $\alpha$-effect.
Such an $\alpha$-effect is susceptible to suppression as it results in oppositely signed 
$\alpha_m$ due to the volume generation (first) term in \Eq{eq: helicity evolution full}.

Interestingly, the NV flux and its divergence are nonzero
even in the absence of large-scale magnetic fields; however it crucially depends on having non-zero
small-scale magnetic fields. These small-scale fields could grow due to the action of a fluctuation dynamo
in a turbulent flow over eddy turn over time scales, even if the random motions are non-helical 
\citep{kazantsev1968,HBD04,Scheko04,BS13}. Moreover, random magnetic fields can also arise due to the magneto rotational 
instability (MRI) in systems like accretion disks \citep{BH98}.

Consider a system like a disk galaxy or an accretion disk, 
rotating with angular velocity $\Omega(r)$, i.e $\meanvec{V} = r\Omega(r) \unitvec{\phi}$.
Its vorticity can be written as $\nabla  \times \meanvec{V} = \unitvec{z} (2\Omega + r d\Omega/dr) = \unitvec{z} \Omega (2-q)$ for $\Omega \propto r^{-q}$.
We then find that $\vec{F}^{NV}$ is predominantly in the $z$ direction.
Different parts of $\vec{\nabla}\cdot\vec{F}^{NV}$ can add or cancel depending on the
the sign of gradients of $\meanBr{v^2}$ and $\meanBr{b^2}$, with the overall sign and magnitude
depending on their relative importance.
Direct simulations suggest that $\meanBr{v^2}$ increases while $\meanBr{b^2}$ decreases away from the disk mid plane
in both disk galaxies (Abhijit Bendre, Private communication using data from \citet{Bendre2015}) and accretion disks (Prasun Dhang, private communication).
Then, \Eq{eq: NVF flux rewrite div} gives
\begin{equation}
\frac{\partial h^b}{\partial t} \approx \frac{ \meanBr{b^2}\Omega \left(2-q \right) }{H}\left[(v_A\tau)^2 C_1 +\meanBr{v^2}\tau^2(C_3-C_2) +C_4\lambda^2 \right].
\label{eq: hb estimate}
\end{equation}
Here, we have defined the Alfv\'en velocity $v_A \defn \left(\meanBr{b^2}/4\pi\rho\right)^{1/2}$, and estimated the gradient as division by the disk scale height $H$ with the sign appropriate for $z>0$.
Let us consider the cases of disk galaxies and accretion disks in turn. 

\subsubsection{Application to disk galaxies}

In the context of disk galaxies, supernovae drive turbulence in the inter-stellar medium, and 
random magnetic fields get generated due to the fluctuation dynamo. 
Current simulations at modest values of the magnetic Reynolds number $R_m \sim 10^3$ suggest 
$v_A \sim 0.5-0.7 v$ at saturation, with $v\defn \sqrt{\meanBr{v^2}}$ 
\citep[c.f.][]{HBD04,Scheko04,Cho+2009,BS13,Sur14,BSB16,F16,Seta20,SF21}, while 
$v_A \sim v$ is obtained by \citet{Eyink13}.\footnote{
See \url{http://turbulence.pha.jhu.edu/Forced_MHD_turbulence.aspx}
}
Observations indicate a random magnetic field a few times larger than the mean field \citep{BeChElBl19,ss21},
but it is not clear which of these should be of order the equipartition value.
For our estimates of the RHS of \Eq{eq: hb estimate}, we adopt $v_A = v$.
We also take $\lambda=v\tau= l$, where $l$ is the eddy scale; and $q=1$, corresponding to a flat rotation curve.
Then $\partial h^b/\partial t \approx 0.076 \meanBr{b^2} \alpha_0$, where
$\alpha_0=(\Omega l^2/H)$ is the standard estimate for the $\alpha$-effect in disk galaxies 
\citep[cf.][]{ss21}. Assuming that the random
field has saturated, neglecting other flux terms, and integrating to get $h^b$ and hence $h^c=h^b/\lambda^2$, 
this gives $\alpha_m = (\tau/3) h^c/(4\pi\rho) \approx 0.025 (t/\tau) \alpha_0$. 
Thus, the NV flux alone can build up a significant magnetic 
$\alpha$-effect in about $40\tau \sim \SI{4e8}{yr}$, where we have taken $\tau \sim \SI{e7}{yr}$ for galactic turbulence. 

In order to obtain more reliable estimates of $\alpha_m$ and its consequences for the growth of the galactic large-scale field, one needs a dynamical quenching model, solving the helicity evolution equation including other fluxes, the mean field equation, and an equation for the the small-scale field simultaneously.
There are also uncertainties associated with the approximations that have been employed to simplify the triple correlation flux.
Nevertheless, the above estimate makes this a promising mechanism for large-scale field generation just due to the influence of helicity fluxes.

\subsubsection{Application to accretion disks}

For MRI-driven turbulence in an accretion disk \citep{BH98}, we expect the random magnetic field to grow on a time scale $\sim \Omega^{-1}$, and saturate with $v_A$ a fraction, say $f_A$ of the sound speed $c_s$, that is $v_A \sim f_A c_s$.
The correlation time is expected to be $\tau \sim 1/\Omega$ and thus $v_A\tau \sim f_A c_s/\Omega \sim f_A H$, where we have used the relation $H \sim c_s/\Omega$ for such disks.
Numerical simulations also show that the magnetic energy dominates the kinetic energy 
by a factor of order 2, is correlated with a scale $\lambda \sim f_\lambda H$ 
and  decreases with $\abs{z}$ (Prasun Dhang; Private communication). 
Adopting $f_A=0.2$, $f_\lambda=0.5$ it turns out that the $C_4$ term dominates the
magnetic helicity generation (with partial cancellation by other terms). 
Taking $q=3/2$ appropriate for
accretion disks and integrating \Eq{eq: hb estimate},
we get $h^c \sim -0.07 (\meanBr{b^2}/H) \Omega t$ 
leading to
an estimate $\alpha_m \approx -0.023 f_A^2 c_s (\Omega t) 
\sim -10^{-3} (\Omega t) \Omega H$.

Moreover, $v_A$ is seen to increase with height in simulations, and so the positive
contributions due to the $C_1$, $C_2$ and $C_3$ terms gain in importance at larger heights.
For example taking now a larger $f_A=0.5=f_\lambda$, $v^2\sim v_A^2/2$ in \Eq{eq: hb estimate}, we now
get $\alpha_m \sim 1.4 \times 10^{-3} (\Omega t) \Omega H$.
These estimates again
need to be firmed up in a detailed calculation, but 
for $\Omega t\sim 1-10$,
look promising to explain both the magnitude
and the sign of the $\alpha$ profile 
in simulations of MRI which also
lead to a large-scale dynamo \citep{Bran95,BranSok02,Davis+2010,GP15,Hogg_2018,Dhang2019,Dhang+2020}. 

\section{Conclusions}
\label{section: conclusions}

We have presented a detailed calculation of several potentially important types of magnetic helicity fluxes  
which can arise from triple correlators in the helicity density evolution equation,
in the presence of random magnetic and velocity fields. 
For this we have assumed 
the velocity field to be incompressible, the random fields to be weakly inhomogeneous and used a quasinormal 
closure approximation with strong cumulant-induced damping.
To begin with, we have been able to obtain a diffusive flux of the magnetic helicity from these hitherto unsimplified terms involving triple correlators 
of the fluctuating fields.
Such a flux has indeed been measured in direct simulations \citep{MitraDiffusiveSim}, and proves useful in dynamical-quenching models
of the large-scale dynamo, which attempt to solve the catastrophic dynamo quenching problem.
We find that the diffusive flux is always accompanied by an advective flux due to gradients in the random velocity and magnetic fields.

A helicity flux similar 
to that first suggested by \citet{newVishniacFlux} has also been obtained from the triple correlators.
This flux arises even in the absence of a large-scale magnetic field, and depends only on the 
random magnetic and velocity fields apart from the mean flow.
Its divergence depends on the gradients of the random fields predominantly along the large-scale vorticity.
It can lead, by itself, to a turbulent EMF and to parity odd magnetic-alpha effect, by merely redistributing
small-scale magnetic helicity within the system.
Given that such a flux requires only rotation and stratification of random fields, it could be important 
to generate large-scale magnetic fields in disk galaxies, accretion disks, or stars. 
More detailed calculations solving dynamical quenching models including the NV flux are required to firm up our conclusions.
Moreover, direct numerical simulations of such stratified rotating systems are important, both to measure such helicity flux contributions and to confirm if they can indeed lead to large-scale dynamo action as envisaged here.

\begin{acknowledgments}
KS thanks Ethan Vishniac for a discussion (a decade back) about his new flux which partly motivated this work. 
We thank Abhijit Bendre and Prasun Dhang for sharing several results from their simulations and for very useful discussions.
\end{acknowledgments}

\appendix

\section{Double-correlators in a weakly inhomogeneous system}
\label{appendix: double correlators}
\subsection{Roberts-Soward expressions}
\label{appendix: RobertsSoward expressions}

We define the Fourier transform as
\begin{equation}
	f(\vec{k}) \defn \int \frac{\d\vec{x}}{(2\pi)^3} \, e^{-i\vec{k}\cdot\vec{x}} f(\vec{x})
	\,.
\end{equation}
To study a two-point correlator $\meanBr{f(\vec{x}^{(1)}) f(\vec{x}^{(2)}) }$, we define $\vec{R} \defn \left( \vec{x}^{(1)} + \vec{x}^{(2)} \right)/2$ and $\vec{r} \defn \vec{x}^{(1)} - \vec{x}^{(2)}$.
In homogeneous turbulence, the correlator would be independent of $\vec{R}$.
Instead, we assume that the turbulence is `weakly inhomogeneous', i.e. that the correlator varies with $\vec{R}$ much more slowly than 
with $\vec{r}$.
Taking a Fourier transform ($\vec{x}^{(1)}, \vec{x}^{(2)} \to \vec{k}^{(1)}, \vec{k}^{(2)}$), we find that $\vec{R} \to \vec{K} = \vec{k}^{(1)} + \vec{k}^{(2)}$ while $\vec{r} \to \vec{k} = \left( \vec{k}^{(1)} - \vec{k}^{(2)} \right)/2$.
Since the turbulence is weakly inhomogeneous, we expand the correlator as a Taylor series in $\vec{K}$, assume the lowest-order terms are isotropic, and discard $\bigO(K^2)$ terms.
Since $\vec{v}$ and $\vec{b}$ are divergenceless,
their double correlators are given by
\citep{robertsSoward}
\begin{align}
	\begin{split}
		\meanBr{ b_i(\tfrac{1}{2}\vec{K}+\vec{k})b_j(\tfrac{1}{2}\vec{K}-\vec{k})} 
		={}& 
		\op{P}_{ij}(\vec{k})M(k,\vec{K}) - \frac{i}{k^2} \eps_{ijc} k_c N(k,\vec{K})
		- \frac{1}{2k^2}\left( k_iK_jM(k,\vec{K}) - k_jK_iM(k,\vec{K}) \right)
		\\& + \frac{i}{2k^4}\left( k_i\eps_{jcd}+k_j\eps_{icd} \right) k_cK_dN(k,\vec{K})
		\,,
	\end{split}\label{eq: bij fourier}
	\\
	\begin{split}
		\meanBr{ v_i(\tfrac{1}{2}\vec{K}+\vec{k})v_j(\tfrac{1}{2}\vec{K}-\vec{k})} 
		={}& 
		\op{P}_{ij}(\vec{k})E(k,\vec{K}) - \frac{i}{k^2}\eps_{ijc}k_cF(k,\vec{K}) 
		- \frac{1}{2k^2}\left(k_iK_jE(k,\vec{K}) - k_jK_iE(k,\vec{K})\right)
		\\& + \frac{i}{2k^4}\left(k_i\eps_{jcd}+k_j\eps_{icd}\right)k_cK_dF(k,\vec{K})
		\,,
	\end{split}\label{eq: vij fourier}
\end{align}
where $4\pi k^2E(\vec{k},\vec{R})$, $8\pi k^2F(\vec{k},\vec{R})$, $k^2M(\vec{k},\vec{R})$, $8\pi N(\vec{k},\vec{R})$ are the kinetic energy, kinetic helicity, magnetic energy and magnetic helicity spectra respectively.
The current helicity spectrum is $8\pi k^2N(\vec{k},\vec{R})$, while $\meanBr{j^2} = \int \d k\, 8\pi k^4 M(k,\vec{R})$.
Note that the double correlators are symmetric under the simultaneous interchange of $i\exchangeArrow j$ and $\vec{k} \to - \vec{k}$,
which is not satisfied by a term in the corresponding expressions of \cite{radler2003,KanduPhysReview}.

\subsection{Correlators in real space}
\label{appendix: double correlators real-space expressions}
To evaluate correlators of interest in real space, we use \Eqs{eq: bij fourier}{eq: vij fourier} along with the following angular integrals:
\begin{align}
	\int\d\Omega \, k_i ={}& 0,
	\quad
	\int\d\Omega \, k_i k_j ={} \frac{4\pi k^2}{3} \delta_{ij},
	\quad
	\int\d\Omega \, k_i k_j k_m ={} 0,
	\quad
	\int\d\Omega \, k_i k_j k_m k_n ={} \frac{4\pi k^4}{15} \left( \delta_{ij}\delta_{mn} + \delta_{im}\delta_{jn} + \delta_{in}\delta_{jm} \right)
	,
\end{align}
where $\d\Omega$ is the angular part of the integral over $\vec{k}$, such that $\d\vec{k} = k^2 \, \d\Omega \, \d k$.
We also use the fact that $a_i(\vec{k}) = \eps_{ijk} \, ik_j \, b_k(\vec{k})/k^2$.
We then find
\begin{align}
	\begin{split}
		\meanBr{v_iv_j }
		={}&
		\frac{1}{3}\delta_{ji}\meanBr{ v^2 }
		+ \bigO(\dh^2)
		\label{eq: vivj correlator}
	\end{split}
	\\
	\begin{split}
		\meanBr{v_i\dh_kv_j }
		={}&
		\frac{1}{6}\left(\delta_{ij}\dh_k\meanBr{v^2 } + \frac{1}{2}\delta_{ik}\dh_j\meanBr{v^2 } - \frac{1}{2}\delta_{jk}\dh_i\meanBr{v^2 } - h^v\eps_{ijk}\right)
		+ \bigO(\dh^2)
		\label{eq: vidkvj correlator}
	\end{split}
	\\
	\begin{split}
		\meanBr{b_ib_j }
		={}&
		\frac{1}{3}\delta_{ij}\meanBr{b^2 }
		+ \bigO(\dh^2)
		\label{eq: bibj correlator}
	\end{split}
	\\
	\begin{split}
		\meanBr{b_i\dh_kb_j }
		={}&
		\frac{1}{6}\left(\delta_{ij}\dh_k\meanBr{b^2 } + \frac{1}{2}\delta_{ik}\dh_j\meanBr{b^2 } - \frac{1}{2}\delta_{jk}\dh_i\meanBr{b^2 } - h^c\eps_{ijk}\right)
		+ \bigO(\dh^2)
		\label{eq: bidkbj correlator}
	\end{split}
	\\
	\begin{split}
		\meanBr{a_ib_j }
		={}&
		\frac{1}{3}\delta_{ij}h^b + \frac{1}{12}\eps_{ijm}\dh_m\meanBr{a^2 }
		+ \bigO(\dh^2) 
		\label{eq: aibj correlator}
	\end{split}
	\\
	\begin{split}
		\meanBr{\left(\dh_jb_i\right)a_k }
		={}&
		\delta_{ik} \frac{7}{30} \dh_j h^b + \delta_{jk} \frac{1}{15} \dh_i h^b - \frac{1}{10} \delta_{ij} \dh_k h^b + \eps_{kji} \frac{1}{6} \meanBr{b^2}
		+ \bigO(\dh^2)
		\label{eq: djbiak correlator}
	\end{split}
	\\
	\begin{split}
		\meanBr{\left(\dh_q\dh_jb_i\right)a_k }
		={}&
		\frac{7}{60} \eps_{kqi} \dh_j\meanBr{b^2}
		+ \frac{7}{60} \eps_{kji} \dh_q\meanBr{b^2}
		- \frac{1}{60} \eps_{kli} \delta_{qj} \dh_l \meanBr{b^2}
		+ \frac{1}{30} \left( \eps_{klj}\delta_{iq} +\eps_{klq}\delta_{ij} \right) \dh_l \meanBr{b^2}
		\\& + \frac{1}{30} \left( \delta_{qk}\delta_{ij} + \delta_{jk}\delta_{iq} - 4\delta_{ik}\delta_{qj} \right) h^c
		+ \bigO(\dh^2)
		\label{eq: dqdjbiak correlator}
	\end{split}
	\\
	\begin{split}
		\meanBr{b_i \dh_j\dh_kb_l}
		={}&
		- \frac{1}{12} \dh_k \eps_{ilj} h^c
		- \frac{1}{12} \dh_j \eps_{ilk} h^c
		- \frac{1}{30} \left( 4\delta_{jk}\delta_{il} - \delta_{ji}\delta_{kl} - \delta_{jl}\delta_{ki} \right) \meanBr{j^2}
		\\& - \frac{1}{30} \left(\delta_{ji}\eps_{lkd} + \delta_{ki}\eps_{ljd} +\delta_{jl}\eps_{ikd} + \delta_{kl}\eps_{ijd} \right) \frac{1}{2}  \dh_d h^c
		+ \bigO(\dh^2)
		\label{eq: bidjdkbl correlator}
	\end{split}
	\\
	\begin{split}
		\meanBr{v_i \dh_j\dh_k v_l } 
		={}&
		- \frac{1}{12} \dh_k \eps_{ilj} h^v
		- \frac{1}{12} \dh_j \eps_{ilk} h^v
		- \frac{1}{30} \left( 4\delta_{jk}\delta_{il} - \delta_{ji}\delta_{kl} - \delta_{jl}\delta_{ki} \right) \meanBr{\omega^2}
		\\& - \frac{1}{30} \left(\delta_{ji}\eps_{lkd} + \delta_{ki}\eps_{ljd} +\delta_{jl}\eps_{ikd} + \delta_{kl}\eps_{ijd} \right) \frac{1}{2}  \dh_d h^v
		+ \bigO(\dh^2)
		\label{eq: vidjdkvl correlator}
	\end{split}
	\\
	\begin{split}
		\meanBr{b_i \dh_e\dh_j\dh_kb_l}
		={}&
		\frac{1}{30} \left(\delta_{ej}\eps_{ilk} + \delta_{ek}\eps_{ilj} + \delta_{jk}\eps_{ile} \right) 
		\int\d k \, 8\pi k^4 N(k,\vec{R})
		+ \bigO(\dh)
		 \label{eq: bdddb correlator}
	\end{split}
	\\
	\begin{split}
		\meanBr{b_i \dh_e\nabla^2b_l}
		={}&
		\frac{1}{6} \eps_{ile} \int\d k\, 8 \pi k^4 N(k,\vec{R})
		- \left( \frac{3}{10}\delta_{il} \dh_e - \frac{7}{60}\delta_{el} \dh_i - \frac{1}{20}\delta_{ei} \dh_l \right) \meanBr{j^2}
		+ \bigO(\dh^2)
		 \label{eq: bdddb correlator with 1 derivative}
	\end{split}
	\\
	\begin{split}
		\meanBr{v_i \dh_e \nabla^2 v_l}
		={}&
		\frac{1}{6} \eps_{ile} \int\d k \, 8\pi k^4 F(k,\vec{R})
		- \left( \frac{3}{10}\delta_{il} \dh_e - \frac{7}{60}\delta_{el} \dh_i - \frac{1}{20}\delta_{ei} \dh_l \right) \meanBr{\omega^2}
		+ \bigO(\dh^2)
		\label{eq: vdddv correlator} 
	\end{split}
	\\
	\begin{split}
		\meanBr{ a_i \left(\dh_d\dh_e\dh_b b_i\right) }
		={}&
		0
		+\bigO(\dh)
	\end{split}
\end{align}
where $\bigO(\dh^n)$ denotes that we have neglected terms with more than $n$ large-scale derivatives.

\section{The triple-correlators}
\label{appendix: triple correlator evaluation}
Using \Eqsn{} \eqref{eq: NS random}, \eqref{eq: induction random in terms of e}, and \eqref{eq: random vector potential time evolution}, we write
\begin{align}
	\begin{split}
		\frac{\dh}{\dh t}\meanBr{ v_ab_ba_c } =&
		- \mean{V}_d \meanBr{ \left(\dh_dv_a\right) b_ba_c }
		- \left(\dh_d\mean{V}_a\right) \meanBr{ v_db_ba_c }
		- \meanBr{  v_d\left(\dh_dv_a\right) b_ba_c  }
		+ \meanBr{ v_d\left(\dh_dv_a\right) } \meanBr{ b_ba_c }
		- \meanBr{ \frac{\dh_ap'}{\rho} b_ba_c  }
		\\& + \frac{\mean{B}_d}{4\pi\rho} \meanBr{ \left(\dh_db_a\right)b_ba_c }
		+ \frac{\dh_d\mean{B}_a}{4\pi\rho} \meanBr{ b_db_ba_c }
		+ \meanBr{  \frac{b_d \left(\dh_db_a\right)}{4\pi\rho} b_ba_c  }
		- \frac{\meanBr{ b_d\left(\dh_db_a\right) }}{4\pi\rho} \meanBr{ b_ba_c }
		\\& + \nu \meanBr{ \left(\dh_d\dh_dv_a\right) b_ba_c }
		+ \left(\dh_d\mean{V}_b\right) \meanBr{ v_ab_da_c }
		- \mean{V}_d \meanBr{ v_a\left(\dh_db_b\right)a_c }
		+ \mean{B}_d \meanBr{ v_a\left(\dh_dv_b\right)a_c }
		- \left(\dh_d\mean{B}_b\right) \meanBr{ v_av_da_c }
		\\& + \meanBr{ v_ab_d\left(\dh_dv_b\right)a_c }
		- \meanBr{ v_av_d\left(\dh_db_b\right)a_c }
		- \eps_{bde}\left(\dh_d\emf_e\right) \meanBr{ v_aa_c }
		+ \eta\meanBr{ v_a\left(\dh_d\dh_db_b\right)a_c }
		+ \eps_{cde} \mean{V}_d \meanBr{ v_ab_bb_e }
		\\& + \eps_{cde} \mean{B}_e \meanBr{ v_ab_bv_d }
		+ \eps_{cde} \meanBr{ v_ab_bv_db_e } 
		- \emf_c \meanBr{ v_ab_b  }
		- \eta\eps_{cde} \meanBr{ v_ab_b\dh_db_e }
		+ \meanBr{ v_ab_b\left(\dh_c\varphi\right) }
		\,.
	\end{split}
\end{align}
In what follows, we drop terms dependent on $\meanvec{B}$, $\nu$, $\eta$, or mixed correlators of the form $\meanBr{ vb }$ or $\meanBr{ va }$.
Assuming the fourth-order correlators above can be expressed as products of second-order correlators along with a damping term, we write
\begin{align}
	\begin{split}
		\frac{\dh}{\dh t}\meanBr{ v_ab_ba_c } =&
		- \mean{V}_d \meanBr{ \left(\dh_dv_a\right) b_ba_c }
		- \left(\dh_d\mean{V}_a\right) \meanBr{ v_db_ba_c }
		- \meanBr{ \frac{\dh_ap'}{\rho} b_ba_c  }
		+ \meanBr{  b_db_b }\meanBr{ \frac{ \left(\dh_db_a\right)}{4\pi\rho} a_c  }
		\\& + \meanBr{  b_da_c }\meanBr{ \frac{ \left(\dh_db_a\right)}{4\pi\rho} b_b  }
		+ \left(\dh_d\mean{V}_b\right) \meanBr{ v_ab_da_c }
		- \mean{V}_d \meanBr{ v_a\left(\dh_db_b\right)a_c }
		+ \meanBr{ v_a\left(\dh_dv_b\right) }\meanBr{ b_da_c }
		\\& - \meanBr{ v_av_d }\meanBr{ \left(\dh_db_b\right)a_c }
		+ \eps_{cde} \mean{V}_d \meanBr{ v_ab_bb_e }
		+ \eps_{cde} \meanBr{ v_av_d }\meanBr{ b_bb_e }
		+ \meanBr{ v_ab_b\left(\dh_c\varphi\right) }
		- \frac{1}{\tau} \meanBr{ v_ab_ba_c }
		.
	\end{split} \label{eq: ddt triple vba}
\end{align}
In steady state,
\begin{align}
	\begin{split}
		\meanBr{ v_ab_ba_c } =
		\tau \bigg[ &
		- \mean{V}_d \meanBr{ \left(\dh_dv_a\right) b_ba_c }
		- \left(\dh_d\mean{V}_a\right) \meanBr{ v_db_ba_c }
		- \meanBr{ \frac{\dh_ap'}{\rho} b_ba_c  }
		+ \meanBr{  b_db_b }\meanBr{ \frac{ \left(\dh_db_a\right)}{4\pi\rho} a_c  }
		\\& + \meanBr{  b_da_c }\meanBr{ \frac{ \left(\dh_db_a\right)}{4\pi\rho} b_b  }
		+  \left(\dh_d\mean{V}_b\right) \meanBr{ v_ab_da_c }
		-  \mean{V}_d \meanBr{ v_a\left(\dh_db_b\right)a_c }
		+  \meanBr{ v_a\left(\dh_dv_b\right) }\meanBr{ b_da_c }
		\\& - \meanBr{ v_av_d }\meanBr{ \left(\dh_db_b\right)a_c }
		+  \eps_{cde} \mean{V}_d \meanBr{ v_ab_bb_e }
		+  \eps_{cde} \meanBr{ v_av_d }\meanBr{ b_bb_e }
		+  \meanBr{ v_ab_b\left(\dh_c\varphi\right) }
		\bigg]
		.
	\end{split} \label{eq: vba before subbing vbb}
\end{align}

Note that above, $\meanBr{ v_ab_ba_c }$ appears on both the LHS and the RHS.
Moreover, the RHS contains other triple correlators such as $\meanBr{ v_ab_bb_e }$.
We can obtain a series for $\meanBr{ v_ab_ba_c }$ by the method of successive approximations, thereby obtaining a series in $\tau$.
For our purposes, we can discard $\bigO(\tau^3)$ contributions to $\meanBr{ v_ab_ba_c }$.

Following exactly the same procedure, we can write
\begin{align}
	\begin{split}
		\meanBr{ v_ab_bb_c } =
		\tau\bigg\{ &
		- \meanBr{ \frac{\dh_ap'}{\rho} b_bb_c }
		+ \frac{1}{4\pi\rho} \meanBr{ b_db_b }\meanBr{ \left(\dh_db_a\right)b_c }
		+ \frac{1}{4\pi\rho} \meanBr{ b_db_c }\meanBr{ \left(\dh_db_a\right)b_b }
		+ \meanBr{ v_a\left(\dh_dv_b\right) }\meanBr{ b_db_c }
		\\& - \meanBr{ v_av_d }\meanBr{ \left(\dh_db_b\right)b_c }
		+ \meanBr{ v_a\left(\dh_dv_c\right) }\meanBr{ b_bb_d }
		- \meanBr{ v_av_d }\meanBr{ b_b\left(\dh_db_c\right)  }
		\bigg\}
		+ \bigO(\tau^2)
	\end{split} \label{eq: vbb triple}
\end{align}
\begin{align}
	\begin{split}
		\meanBr{ \left(\dh_dv_a\right)b_ba_c }
		={}&
		\tau \bigg[
		- \meanBr{ \frac{\dh_a\dh_dp'}{\rho} b_ba_c }
		+ \frac{1}{4\pi\rho} \meanBr{ \left( \dh_db_j \right) b_b } \meanBr{ \left( \dh_jb_a \right) a_c }
		+ \frac{1}{4\pi\rho} \meanBr{ \left( \dh_db_j \right) a_c } \meanBr{ \left( \dh_jb_a \right) b_b }
		\\&\continuedTerm + \frac{1}{4\pi\rho} \meanBr{ b_j b_b } \meanBr{ \left( \dh_d\dh_jb_a \right) a_c }
		+ \frac{1}{4\pi\rho} \meanBr{ b_j a_c } \meanBr{ \left( \dh_d\dh_jb_a \right) b_b }
		+ \meanBr{ \left(\dh_dv_a\right) \left(\dh_jv_b\right) } \meanBr{ b_j  a_c }
		\\&\continuedTerm - \meanBr{ \left(\dh_dv_a\right) v_j } \meanBr{ \left(\dh_jb_b\right) a_c }
		+ \eps_{clm} \meanBr{ \left(\dh_dv_a\right) v_l } \meanBr{ b_b  b_m }
		+ \meanBr{ \left(\dh_dv_a\right)b_b \left( \dh_c\varphi \right) }
		\bigg]
		+ \bigO(\tau^2)
	\end{split} \label{eq: dvba correlator horrible}
\end{align}
\begin{align}
	\begin{split}
		\meanBr{ v_a\left(\dh_db_b\right)a_c }
		={}&
		\tau\bigg[
		- \meanBr{ \frac{\dh_ap'}{\rho} \left(\dh_db_b\right)a_c }
		+ \frac{1}{4\pi\rho} \meanBr{ b_j \left(\dh_db_b\right) } \meanBr{ \left( \dh_jb_a \right) a_c }
		+ \frac{1}{4\pi\rho} \meanBr{ b_j a_c } \meanBr{ \left( \dh_jb_a \right) \left(\dh_db_b\right) }
		\\&\continuedTerm + \meanBr{ v_a  \left( \dh_jv_b \right) } \meanBr{ \left( \dh_d b_j \right) a_c }
		+ \meanBr{ v_a \left( \dh_d\dh_jv_b \right) } \meanBr{ b_j a_c }
		- \meanBr{ v_a \left( \dh_d v_j \right) } \meanBr{ \left( \dh_jb_b \right) a_c }
		\\&\continuedTerm - \meanBr{ v_a v_j } \meanBr{ \left( \dh_d\dh_jb_b \right) a_c }
		+ \eps_{clm} \meanBr{ v_a v_l } \meanBr{ \left(\dh_db_b\right) b_m }
		+ \meanBr{ v_a\left(\dh_db_b\right) \dh_c\varphi }
		\bigg]
		+ \bigO(\tau^2)
	\end{split} \label{eq: vdba correlator horrible}
\end{align}

In appendix \ref{appendix: double correlators real-space expressions}, we have listed expressions for the various double-correlators appearing in \Eqsn{} \eqref{eq: vba before subbing vbb}, \eqref{eq: vbb triple}, \eqref{eq: dvba correlator horrible}, and \eqref{eq: vdba correlator horrible}, assuming weakly inhomogeneous turbulence.
Using these, discarding terms with more than one large-scale derivative, and substituting \Eqsn{} \eqref{eq: vbb triple}, \eqref{eq: dvba correlator horrible}, and \eqref{eq: vdba correlator horrible} into \Eq{eq: vba before subbing vbb}, a lengthy but straightforward calculation gives us the terms $I^\text{triple}_1$ and $I^\text{triple}_2$ (which appear in the evolution equation for the helicity density, \Eq{eq: helicity evolution}):
\begin{align*}
		I^\text{triple}_1 
		={}& \dh_j\bigg[
		- \tau \meanBr{ \frac{\dh_ip'}{\rho} b_ja_i  }
		+ \tau \frac{2}{9} \frac{\meanBr{b^2}}{4\pi\rho} \dh_j h^b
		- \tau \frac{1}{36} \frac{1}{4\pi\rho} h^c \dh_j\meanBr{a^2}
		\\&\continuedTerm + \tau \frac{1}{9} h^b \dh_j\meanBr{ v^2 } 
		+ \tau \frac{1}{36} h^v \dh_j\meanBr{a^2}
		- \tau \frac{1}{9}\meanBr{v^2} \dh_j h^b
		+ \tau \meanBr{ v_ib_j\left(\dh_i\varphi\right) }
		\\&\continuedTerm + \mean{V}_d \tau^2 \bigg\{ 
		\meanBr{ a_i \dh_d \left( \frac{\dh_i p'}{\rho} b_j \right) }
		+ \eps_{jdi} \frac{1}{36} \frac{h^c}{4\pi\rho} \dh_i h^b
		- \eps_{jdi} \frac{1}{24} \frac{1}{4\pi\rho} \meanBr{b^2} \dh_i\meanBr{ b^2 }
		\\&\continuedTerm\continuedTerm + \delta_{dj} \frac{1}{18} \frac{1}{4\pi\rho} \meanBr{b^2}  h^c 
		+ \eps_{jdi} \frac{11}{72} \meanBr{b^2} \dh_i\meanBr{ v^2 } 
		- \delta_{jd} \frac{1}{9} \meanBr{b^2} h^v
		- \meanBr{ \left( \dh_i\varphi \right) \dh_d\left( v_ib_j \right)  }
		\\&\continuedTerm\continuedTerm - \eps_{jdi} \frac{1}{36} h^v \dh_i h^b 
		+ \delta_{jd} \frac{1}{18} h^v \meanBr{b^2}
		+ \eps_{jdi} \frac{1}{18} \meanBr{v^2} \dh_i\meanBr{b^2}
		- \eps_{ide} \meanBr{ \frac{\dh_ip'}{\rho} b_jb_e }
		\bigg\}
		\\&\continuedTerm - \left(\dh_d\mean{V}_i\right) \tau^2 \bigg\{
		- \meanBr{ \frac{\dh_dp'}{\rho} b_ja_i  }
		+ \epsilon_{ijd} \frac{1}{18} \frac{1}{4\pi\rho} \meanBr{b^2} \meanBr{b^2}
		+ \epsilon_{jid} \frac{1}{18} \frac{1}{4\pi\rho} h^b  h^c 
		\\&\continuedTerm\continuedTerm + \epsilon_{dij} \frac{1}{18} h^v h^b 
		+ \eps_{idj} \frac{1}{18}\meanBr{v^2} \meanBr{b^2}
		+ \meanBr{ v_db_j\left(\dh_i\varphi\right) }
		\bigg\}
		\\&\continuedTerm + \left(\dh_d\mean{V}_j\right) \tau^2 \bigg\{
		- \meanBr{ \frac{\dh_ip'}{\rho} b_da_i  }
		+ \meanBr{ v_ib_d\left(\dh_i\varphi\right) }
		\bigg\} 
		\bigg] 
		+ \bigO(\tau^3)
		\,,
		\taghere
		\label{eq: I1 final}
\end{align*}
and
\begin{align*}
		I^\text{triple}_2 
		={}& - \dh_j\bigg[
		- \tau \meanBr{ \frac{\dh_jp'}{\rho} b_ia_i  }
		+ \tau \frac{1}{9} \frac{\meanBr{b^2}}{4\pi\rho} \dh_j h^b
		+ \tau \frac{1}{9} \frac{1}{4\pi\rho} h^b \dh_j\meanBr{ b^2 }
		\\&\continuedTerm + \tau \frac{1}{36} \frac{1}{4\pi\rho} h^c \dh_j\meanBr{a^2}
		- \tau \frac{1}{36} h^v \dh_j\meanBr{a^2}
		- \tau \frac{2}{9}\meanBr{v^2} \dh_j h^b 
		+ \tau \meanBr{ v_jb_i\left(\dh_i\varphi\right) }
		\\&\continuedTerm + \mean{V}_d \tau^2 \bigg\{ 
		\meanBr{ a_i \dh_d \left( \frac{\dh_j p'}{\rho} b_i \right) }
		- \eps_{jdi} \frac{1}{36} \frac{h^c}{4\pi\rho} \dh_i h^b 
		+ \eps_{jdi} \frac{1}{72} \frac{1}{4\pi\rho} \meanBr{b^2} \dh_i\meanBr{ b^2 } 
		\\&\continuedTerm\continuedTerm + \frac{1}{18} \frac{1}{4\pi\rho} \meanBr{b^2} \delta_{dj} h^c
		- \meanBr{ \left( \dh_i\varphi \right) \dh_d\left( v_jb_i \right)  }
		- \eps_{jdi} \frac{1}{72} \meanBr{b^2} \dh_i\meanBr{ v^2 } 
		+ \eps_{jdi} \frac{1}{36} h^v \dh_i h^b
		\\&\continuedTerm\continuedTerm - \delta_{jd} \frac{1}{18} h^v \meanBr{b^2}
		- \eps_{jdi} \frac{1}{18} \meanBr{v^2} \dh_i\meanBr{ b^2 } 
		- \eps_{ide} \meanBr{ \frac{\dh_jp'}{\rho} b_ib_e }
		\bigg\}
		\\&\continuedTerm - \left(\dh_d\mean{V}_j\right) \tau^2 \bigg\{
		- \meanBr{ \frac{\dh_dp'}{\rho} b_ia_i  }
		+ \meanBr{ v_db_i\left(\dh_i\varphi\right) }
		\bigg\}
		\\&\continuedTerm + \left(\dh_d\mean{V}_i\right) \tau^2 \bigg\{
		- \meanBr{ \frac{\dh_jp'}{\rho} b_da_i  }
		+ \epsilon_{idj} \frac{1}{18} \frac{1}{4\pi\rho} \meanBr{b^2}\meanBr{b^2}
		+ \epsilon_{dij} \frac{1}{18} \frac{1}{4\pi\rho} h^b  h^c 
		\\&\continuedTerm\continuedTerm + \epsilon_{jid} \frac{1}{18} h^v h^b
		+ \eps_{ijd} \frac{1}{18}\meanBr{v^2} \meanBr{b^2}
		+ \meanBr{ v_jb_d\left(\dh_i\varphi\right) }
		\bigg\} 
		\bigg]
		+ \bigO(\tau^3)
		\,.
		\taghere
		\label{eq: I2 final}
\end{align*}

Substituting \Eqs{eq: I1 final}{eq: I2 final} into \Eq{eq: helicity evolution}, we obtain
\begin{align}
	\begin{split}
		\left(\frac{\dh h^b}{\dh t}\right)_{\text{triple}} 
		=
		\dh_j\Bigg[ &
			\overbrace{ 
			\frac{\tau}{9} \left( \meanBr{v^2} + \frac{\meanBr{b^2}}{4\pi\rho} \right) \dh_j h^b 
			}^{\text{diffusion}} {}
			+ \tau \frac{1}{18} \left( h^v - \frac{h^c}{4\pi\rho} \right) \dh_j\meanBr{a^2}
			+ \overbrace{
			\frac{\tau}{9} h^b \dh_j \left( \meanBr{ v^2 } - \frac{\meanBr{ b^2 }}{4\pi\rho} \right)
			}^{\text{random advection}} {}
			\\& - \tau \meanBr{ \frac{\dh_ip'}{\rho} b_ja_i  }
			+ \tau \meanBr{ \frac{\dh_jp'}{\rho} b_ia_i  }
			+ \tau \meanBr{ v_ib_j\left(\dh_i\varphi\right) }
			- \tau \meanBr{ v_jb_i\left(\dh_i\varphi\right) }
			\\& + \mean{V}_d \tau^2 \Bigg\{ 
			\overbrace{ {}
			- \eps_{jdi} \frac{1}{18} \frac{1}{4\pi\rho} \meanBr{b^2} \dh_i\meanBr{ b^2 }
			+ \eps_{jdi} \frac{1}{6} \meanBr{b^2} \dh_i\meanBr{ v^2 } 
			+ \eps_{jdi} \frac{1}{9} \meanBr{v^2} \dh_i\meanBr{b^2}
			}^{\text{NV}}
			\\&\continuedTerm + \eps_{jdi} \frac{1}{18} \frac{h^c}{4\pi\rho} \dh_i h^b
			- \eps_{jdi} \frac{1}{18} h^v \dh_i h^b 
			+ \meanBr{ a_i \dh_d \left( \frac{\dh_i p'}{\rho} b_j \right) }
			- \eps_{ide} \meanBr{ \frac{\dh_ip'}{\rho} b_jb_e }
			\\&\continuedTerm - \meanBr{ a_i \dh_d \left( \frac{\dh_j p'}{\rho} b_i \right) }
			+ \meanBr{ \left( \dh_i\varphi \right) \dh_d\left( v_jb_i \right)  }
			- \meanBr{ \left( \dh_i\varphi \right) \dh_d\left( v_ib_j \right)  }
			+ \eps_{ide} \meanBr{ \frac{\dh_jp'}{\rho} b_ib_e }
			\Bigg\}
			\\& - \left(\dh_d\mean{V}_i\right) \tau^2 \bigg(
			- \meanBr{ \frac{\dh_dp'}{\rho} b_ja_i  }
			+ \meanBr{ v_db_j\left(\dh_i\varphi\right) }
			- \meanBr{ \frac{\dh_jp'}{\rho} b_da_i  }
			+ \meanBr{ v_jb_d\left(\dh_i\varphi\right) }
			\bigg)
			\\& + \left(\dh_d\mean{V}_j\right) \tau^2 \bigg(
			- \meanBr{ \frac{\dh_ip'}{\rho} b_da_i  }
			+ \meanBr{ v_ib_d\left(\dh_i\varphi\right) }
			- \meanBr{ \frac{\dh_dp'}{\rho} b_ia_i  }
			+ \meanBr{ v_db_i\left(\dh_i\varphi\right) }
			\bigg)
		\Bigg]
		\,.
	\end{split}  \label{eq: dhbdt unsimplified pressure phi}
\end{align}

\section{The nonlocal terms}
\label{appendix: nonlocal}
\subsection{Correlators involving the pressure}
\subsubsection{\texorpdfstring{$\meanBr{ \frac{\dh_ap'}{\rho}b_ba_c }$}{<(Dp)ba>}}

\label{appendix: dpba correlator}

If we drop the viscosity and mean magnetic field terms in \Eq{eq: NS random}, take its divergence, and solve for the pressure, we obtain
\begin{align}
	\begin{split}
		p' ={}& \rho\nabla^{-2}\dh_i\left(- \mean{V}_j\dh_jv_i - v_j\dh_j\mean{V}_i - v_j\dh_jv_i + \meanBr{ v_j\dh_jv_i } 
		+ \frac{b_j\dh_jb_i}{4\pi\rho} - \frac{\meanBr{ b_j\dh_jb_i }}{4\pi\rho}\right)
	\end{split} \label{eq: expression for pressure}
\end{align}
where $\nabla^{-2}$ denotes the integral operator that is the inverse of the Laplacian, given by
\begin{equation}
	\nabla^{-2}f(\vec{x}) = -\int\d^3y\frac{1}{4\pi\left|\vec{x}-\vec{y}\right|}f(\vec{y})
	\,.
	\label{eq: inverse Laplacian}
\end{equation}

Using \Eqs{eq: expression for pressure}{eq: inverse Laplacian}, we write
\begin{align}
	\begin{split}
		\meanBr{ \frac{\dh_ap'}{\rho}b_ba_c }
		={}& 
		- \int\d^3y\frac{x_a-y_a}{4\pi\left|\vec{x}-\vec{y}\right|^3}\frac{\dh\mean{V}_e(\vec{y})}{\dh y_d}\meanBr{ b_b(\vec{x})a_c(\vec{x})\frac{\dh v_d(\vec{y})}{\dh y_e} }
		\\& - \int\d^3y\frac{x_a-y_a}{4\pi\left|\vec{x}-\vec{y}\right|^3}\frac{\dh\mean{V}_d(\vec{y})}{\dh y_e}\meanBr{ b_b(\vec{x})a_c(\vec{x})\frac{\dh v_e(\vec{y})}{\dh y_d} }
		\\& - \int\d^3y\frac{x_a-y_a}{4\pi\left|\vec{x}-\vec{y}\right|^3}\meanBr{ b_b(\vec{x})a_c(\vec{x})\frac{\dh v_e(\vec{y})}{\dh y_d}\frac{\dh v_d(\vec{y})}{\dh y_e} }
		\\& + \int\d^3y\frac{x_a-y_a}{4\pi\left|\vec{x}-\vec{y}\right|^3}\meanBr{ b_b(\vec{x})a_c(\vec{x}) }\meanBr{ \frac{\dh v_e(\vec{y})}{\dh y_d}\frac{\dh v_d(\vec{y})}{\dh y_e} }
		\\& + \int\d^3y\frac{x_a-y_a}{4\pi\left|\vec{x}-\vec{y}\right|^3}\frac{1}{4\pi\rho}\meanBr{ b_b(\vec{x})a_c(\vec{x})\frac{\dh b_e(\vec{y})}{\dh y_d}\frac{\dh b_d(\vec{y})}{\dh y_e} }
		\\& - \int\d^3y\frac{x_a-y_a}{4\pi\left|\vec{x}-\vec{y}\right|^3}\frac{1}{4\pi\rho}\meanBr{ b_b(\vec{x})a_c(\vec{x}) }\meanBr{ \frac{\dh b_e(\vec{y})}{\dh y_d}\frac{\dh b_d(\vec{y})}{\dh y_e} }
		.
	\end{split}
\end{align}
Using the quasinormal approximation for the fourth-order correlators and discarding correlators of the form $\meanBr{ vb }$ gives
\begin{align}
	\begin{split}
		\meanBr{ \frac{\dh_ap'}{\rho}b_ba_c }
		={}& 
		- 2\int\d^3y\frac{x_a-y_a}{4\pi\left|\vec{x}-\vec{y}\right|^3}\frac{\dh\mean{V}_e(\vec{y})}{\dh y_d}\frac{\dh}{\dh y_e}\meanBr{ b_b(\vec{x})a_c(\vec{x})v_d(\vec{y}) }
		\\& + 2\int\d^3y\frac{x_a-y_a}{4\pi\left|\vec{x}-\vec{y}\right|^3}\frac{1}{4\pi\rho}\frac{\dh}{\dh y_d}\meanBr{ b_b(\vec{x})b_e(\vec{y}) }\frac{\dh}{\dh y_e}\meanBr{ a_c(\vec{x})b_d(\vec{y}) }
		.
	\end{split} \label{eq: dhpba correlator before VC}
\end{align}
Using the Vishniac-Cho approximation, we write the above as
\begin{align}
	\begin{split}
		\meanBr{ \frac{\dh_ap'}{\rho}b_ba_c }
		\approx{}& 
		\frac{2\lambda^2}{3} \frac{\dh\mean{V}_e(\vec{x})}{\dh x_d} \meanBr{ b_b(\vec{x})a_c(\vec{x}) \frac{\dh^2 v_d(\vec{x})}{\dh x_e \dh x_a}  }
		- \frac{2\lambda^2}{3} \frac{1}{4\pi\rho} \meanBr{ b_b(\vec{x}) \frac{\dh^2 b_e(\vec{x})}{\dh x_d \dh x_a}  } \meanBr{ a_c(\vec{x}) \frac{\dh b_d(\vec{x})}{\dh x_e}  }
		\\& - \frac{2\lambda^2}{3} \frac{1}{4\pi\rho}\meanBr{ b_b(\vec{x}) \frac{\dh b_e(\vec{x})}{\dh x_d}  } \meanBr{ a_c(\vec{x}) \frac{\dh^2 b_d(\vec{x})}{\dh x_e \dh x_a}  }
		.
	\end{split} \label{eq: dpba correlator after VC}
\end{align}
If we substitute \Eqsn{} \eqref{eq: bidkbj correlator}, \eqref{eq: djbiak correlator}, \eqref{eq: dqdjbiak correlator}, and \eqref{eq: bidjdkbl correlator} for the various correlators in the above equation, we obtain
\begin{align}
	\begin{split}
		\meanBr{ \frac{\dh_ap'}{\rho}b_ba_c }
		\approx{}& 
		\frac{2\lambda^2}{3} \frac{\dh\mean{V}_e(\vec{x})}{\dh x_d} \meanBr{ b_b(\vec{x})a_c(\vec{x}) \frac{\dh^2 v_d(\vec{x})}{\dh x_e \dh x_a}  }
		+ \frac{13\lambda^2}{675} \frac{1}{4\pi\rho} \meanBr{j^2} \delta_{ca}\dh_b h^b 
		+ \frac{\lambda^2}{1350} \frac{1}{4\pi\rho} \meanBr{j^2} \delta_{cb} \dh_a h^b
		\\& - \frac{\lambda^2}{150} \frac{1}{4\pi\rho} \meanBr{j^2} \delta_{ab} \dh_c h^b
		+ \frac{17\lambda^2}{270} \frac{1}{4\pi\rho} \delta_{bc}\meanBr{b^2} \dh_a h^c
		- \frac{\lambda^2}{90} \frac{1}{4\pi\rho} \delta_{ac}  \meanBr{b^2} \dh_b h^c 
		\\& + \frac{\lambda^2}{135} \frac{1}{4\pi\rho} \delta_{ba} \meanBr{b^2} \dh_c h^c
		+ \frac{\lambda^2}{54} \frac{1}{4\pi\rho} \eps_{cba} \meanBr{b^2} \meanBr{j^2}
		- \frac{\lambda^2}{54} \frac{1}{4\pi\rho} \eps_{bac} h^c h^c
		\,.
	\end{split} \label{eq: dpba correlator}
\end{align}

Now, the term in \Eq{eq: dpba correlator} containing the mean velocity is required only for the index choices $a=c$ or $b=c$.
Using \Eq{eq: baddv triple correlator a equal c} and then \Eqsn{} \eqref{eq: dqdjbiak correlator}, \eqref{eq: bidjdkbl correlator}, \eqref{eq: vidjdkvl correlator}, \eqref{eq: bdddb correlator}, and \eqref{eq: vdddv correlator}, we write \Eq{eq: dpba correlator} for $a=c$ as
\begin{align}
	\begin{split}
		\meanBr{ \frac{\dh_ip'}{\rho}b_ja_i }
		\approx{}& 
		\frac{2\tau \lambda^2}{3} \left( \dh_d \mean{V}_e \right) \bigg[
		- \frac{1}{4\pi\rho} \frac{1}{36} \epsilon_{ejd} \meanBr{j^2} \meanBr{b^2}
		- \frac{1}{4\pi\rho} \epsilon_{jde} \frac{1}{36} h^c h^c
		+ \frac{1}{4\pi\rho} \frac{1}{18} \eps_{jde} h^b \int\d k \, 8\pi k^4 N(k,\vec{R})
		\\&\continuedTerm - \frac{1}{18} \eps_{jde} h^b \int\d k \, 8\pi k^4 F(k,\vec{R}) 
		- \frac{1}{36} \epsilon_{edj} \meanBr{\omega^2} \meanBr{b^2}
		\bigg]
		\\& + \frac{7\lambda^2}{135} \frac{1}{4\pi\rho} \meanBr{j^2} \dh_j h^b 
		+ \frac{\lambda^2}{27} \frac{1}{4\pi\rho} h^c \dh_j\meanBr{ b^2 } 
		.
	\end{split} \label{eq: dpba correlator with V a equals c}
\end{align}
Similarly, we can use \Eq{eq: baddv triple correlator b equal c} and then \Eqsn{} \eqref{eq: bidjdkbl correlator} and \eqref{eq: dqdjbiak correlator} to write \Eq{eq: dpba correlator} for $b=c$ as
\begin{align}
	\begin{split}
		\meanBr{ \frac{\dh_jp'}{\rho}b_ia_i }
		\approx{}& 
		\frac{2\lambda^2}{135} \frac{1}{4\pi\rho} \meanBr{j^2} \dh_j h^b 
		+ \frac{2\lambda^2}{27} \frac{1}{4\pi\rho} \meanBr{b^2} \dh_j h^c
		+ \frac{\lambda^2}{9} \frac{1}{4\pi\rho} h^c \dh_j\meanBr{ b^2 }
		.
	\end{split} \label{eq: dpba correlator with V b equals c}
\end{align}

\subsubsection{\texorpdfstring{$\meanBr{ \frac{\dh_a p'}{\rho} b_b b_c }$}{<(Dp)bb>}}
This correlator appears in \Eq{eq: dhbdt unsimplified pressure phi} multiplied by a $\tau^2$ factor.
Similar to \Eq{eq: dhpba correlator before VC}, we write
\begin{equation}
		\meanBr{ \frac{\dh_ap'}{\rho}b_bb_c } =
		2\int\d^3y\frac{x_a-y_a}{4\pi\left|\vec{x}-\vec{y}\right|^3} \frac{1}{4\pi\rho}\meanBr{ b_b(\vec{x})\frac{\dh b_e(\vec{y})}{\dh y_d} }\meanBr{ b_c(\vec{x})\frac{\dh b_d(\vec{y})}{\dh y_e} }
		.
\end{equation}
Using the Vishniac-Cho approximation, this becomes
\begin{equation}
		\meanBr{ \frac{\dh_ap'}{\rho}b_bb_c }
		\approx
		- \frac{2\lambda^2}{3} \frac{1}{4\pi\rho} \meanBr{ b_b \dh_d b_e } \meanBr{ b_c \dh_a \dh_e b_d }
		- \frac{2\lambda^2}{3} \frac{1}{4\pi\rho} \meanBr{ b_b \dh_a \dh_d b_e } \meanBr{ b_c \dh_e b_d }
		.
\end{equation}
Examining \Eq{eq: dhbdt unsimplified pressure phi}, we see that the correlator we need is $\epsilon_{ide} \meanBr{ \frac{\dh_ip'}{\rho}b_jb_e }$.
Using \Eqsn{} \eqref{eq: bidkbj correlator} and \eqref{eq: bidjdkbl correlator} along with the above, we get
\begin{equation}
	\epsilon_{ide} \meanBr{ \frac{\dh_ip'}{\rho}b_jb_e } 
	=
	\frac{\lambda^2}{18} \frac{1}{4\pi\rho} \eps_{jid} h^c \dh_i h^c
	\,.
	\label{eq: eps dpbb correlator}
\end{equation}

\subsubsection{\texorpdfstring{$\meanBr{ a_a \dh_d \left( \frac{\dh_b p'}{\rho} b_c \right) }$}{<aD((Dp)b)>}}
Note that this correlator appears in \Eq{eq: dhbdt unsimplified pressure phi} multiplied by $\tau^2$.
We need to keep up to one large-scale derivative in it, but we can discard terms involving the mean velocity.
We write
\begin{equation}
	\meanBr{ a_a \dh_d \left( \frac{\dh_b p'}{\rho} b_c \right) } = 
	 \dh_d \meanBr{ \frac{\dh_b p'}{\rho} b_c a_a } - \meanBr{ \frac{\dh_b p'}{\rho} b_c \left( \dh_d a_a \right) }
\end{equation}
Following steps similar to those in appendix \eqref{appendix: dpba correlator}, the second correlator on the RHS is
\begin{align}
	\begin{split}
		\meanBr{ \frac{\dh_bp'}{\rho}b_c \left(\dh_d a_a\right) }
		\approx{}& 
		- \frac{2\lambda^2}{3} \frac{1}{4\pi\rho} \meanBr{ b_c \left(\dh_f\dh_b b_e\right) } \dh_d \meanBr{ a_a \left(\dh_e b_f\right) }
		+ \frac{2\lambda^2}{3} \frac{1}{4\pi\rho} \meanBr{ b_c \left(\dh_f\dh_b b_e\right) } \meanBr{ a_a \left(\dh_d\dh_e b_f\right) }
		\\& - \frac{2\lambda^2}{3} \frac{1}{4\pi\rho}\meanBr{ b_c \left(\dh_f b_e\right) } \dh_d \meanBr{ a_a \left(\dh_e\dh_b b_f\right) }
		+ \frac{2\lambda^2}{3} \frac{1}{4\pi\rho}\meanBr{ b_c \left(\dh_f b_e\right) } \meanBr{ a_a \left(\dh_d\dh_e\dh_b b_f\right) }
		.
	\end{split}
\end{align}
Using the above along with \Eq{eq: dpba correlator after VC}, we write
\begin{align}
	\begin{split}
		\meanBr{ a_a \dh_d \left( \frac{\dh_b p'}{\rho} b_c \right) }
		={}&
		- \frac{2\lambda^2}{3} \frac{1}{4\pi\rho}  \meanBr{ a_a \left(\dh_e b_f\right) } \dh_d \meanBr{ b_c \left(\dh_f\dh_b b_e\right) }
		- \frac{2\lambda^2}{3} \frac{1}{4\pi\rho} \meanBr{ a_a \left(\dh_e\dh_b b_f\right) } \dh_d \meanBr{ b_c \left(\dh_f b_e\right) }
		\\& - \frac{2\lambda^2}{3} \frac{1}{4\pi\rho} \meanBr{ b_c \left(\dh_f\dh_b b_e\right) } \meanBr{ a_a \left(\dh_d\dh_e b_f\right) }
		- \frac{2\lambda^2}{3} \frac{1}{4\pi\rho}\meanBr{ b_c \left(\dh_f b_e\right) } \meanBr{ a_a \left(\dh_d\dh_e\dh_b b_f\right) }
		.
	\end{split}
\end{align}

Now, we only need the values of the above correlator for $b=a$ and $c=a$.
They are
\begin{equation}
		\meanBr{ a_i \dh_d \left( \frac{\dh_i p'}{\rho} b_c \right) }
		=
		- \frac{2\lambda^2}{3} \frac{1}{4\pi\rho} \frac{4}{30} \meanBr{j^2} \frac{1}{3} \delta_{dc} h^c 
		- \frac{\lambda^2}{270} \frac{1}{4\pi\rho} \meanBr{j^2}  \eps_{idc} \dh_i  \meanBr{b^2}
		- \frac{\lambda^2}{30} \frac{1}{4\pi\rho} h^c \eps_{cdf} \dh_f h^c
		\label{eq: d(dpb)a correlator b equal a}
\end{equation}
and
\begin{equation}
		\meanBr{ a_i \dh_d \left( \frac{\dh_b p'}{\rho} b_i \right) }
		=
		\frac{2\lambda^2}{135} \frac{1}{4\pi\rho} \eps_{idb} \meanBr{j^2} \dh_i \meanBr{b^2}
		- \frac{4\lambda^2}{135} \frac{1}{4\pi\rho} \delta_{db} \meanBr{j^2} h^c
		- \frac{2\lambda^2}{135} \frac{1}{4\pi\rho} h^c \eps_{dbg} \dh_g h^c
		.
		\label{eq: d(dpb)a correlator c equal a}
\end{equation}

\subsection{Correlators involving the scalar potential}
\label{appendix: potential correlators}
Taking the divergence of the evolution equation for the vector potential, \Eq{eq: random vector potential time evolution}, we obtain
\begin{equation}
	\dh_ke_k = \nabla^2\varphi
	\quad\implies\quad
	\varphi = \nabla^{-2}\dh_ke_k
	\,,
\end{equation}
where $\nabla^{-2}$ is the inverse of the Laplacian, given in \Eq{eq: inverse Laplacian}.

On the other hand, taking the divergence of both sides of \Eq{eq: e definition} gives us
\begin{equation}
	\dh_ke_k = -\eps_{klm}\dh_k\left(\mean{V}_lb_m\right) - \eps_{klm}\dh_k\left(v_lb_m\right)
\end{equation}
where we have dropped terms dependent on $\meanvec{B}$ and $\emf$.

The scalar potential is then
\begin{equation}
	\varphi = \nabla^{-2}\left[-\eps_{klm}\dh_k\left(\mean{V}_lb_m\right) - \eps_{klm}\dh_k\left(v_lb_m\right)\right] 
	.
	\label{eq: expression for phi}
\end{equation}

\subsubsection{\texorpdfstring{$\meanBr{ v_ab_b\left(\dh_c\phi\right) }$}{<vbDphi>}}
Using \Eq{eq: expression for phi}, we write
\begin{align}
	\begin{split}
		\meanBr{ v_ab_b\left(\dh_c\phi\right) }
		={}& 
		\eps_{klm}\int\d^3r\frac{r_c}{4\pi\left|\vec{r}\right|^3}\frac{\dh \mean{V}_l(\vec{y}) }{\dh y_k}\meanBr{ v_a(\vec{x})b_b(\vec{x})b_m(\vec{x}+\vec{r})  }
		\\& + \eps_{klm}\int\d^3r\frac{r_c}{4\pi\left|\vec{r}\right|^3} \mean{V}_l(\vec{x}+\vec{r}) \frac{\dh}{\dh r_k}\meanBr{ v_a(\vec{x})b_b(\vec{x})b_m(\vec{x}+\vec{r}) }
		\\& + \eps_{klm}\int\d^3r\frac{r_c}{4\pi\left|\vec{r}\right|^3}\frac{\dh}{\dh r_k}\meanBr{ v_a(\vec{x})b_b(\vec{x}) v_l(\vec{x}+\vec{r})b_m(\vec{x}+\vec{r}) }
		.
	\end{split}
\end{align}
Following a procedure similar to that used to derive \Eq{eq: dhpba correlator before VC} and then using the Vishniac-Cho approximation, we write
\begin{align}
	\begin{split}
		\meanBr{ v_ab_b\left(\dh_c\phi\right) }
		\approx{}& 
		\eps_{klm} \frac{\lambda^2}{3} \left( \dh_k \mean{V}_l \right) \meanBr{ v_a b_b \dh_c b_m  } 
		+ \eps_{klm} \frac{\lambda^2}{3} \mean{V}_l  \meanBr{ v_a b_b \dh_k \dh_c b_m  }
		+ \eps_{klm} \frac{\lambda^2}{3} \left( \dh_c \mean{V}_l \right) \meanBr{ v_a b_b \dh_k b_m  }
		\\& + \eps_{klm} \frac{\lambda^2}{3} \meanBr{  v_a \dh_c \dh_k v_l  } \meanBr{  b_b b_m  } 
		+ \eps_{klm} \frac{\lambda^2}{3} \meanBr{  v_a \dh_c v_l  } \meanBr{  b_b \dh_k b_m  }
		+ \eps_{klm} \frac{\lambda^2}{3} \meanBr{  v_a \dh_k v_l  } \meanBr{  b_b \dh_c b_m  }
		\\& + \eps_{klm} \frac{\lambda^2}{3} \meanBr{  v_a  v_l  } \meanBr{  b_b \dh_c \dh_k b_m  }
		.
	\end{split} \label{eq: vbdphi correlator intermediate 1}
\end{align}
Using \Eqsn{} \eqref{eq: vidjdkvl correlator}, \eqref{eq: bibj correlator}, \eqref{eq: vidkvj correlator}, \eqref{eq: bidkbj correlator}, \eqref{eq: vivj correlator}, and \eqref{eq: bidjdkbl correlator}, we keep up to one large-scale derivative and write the part of the correlator that does not involve $\mean{V}$ as
\begin{align}
	\begin{split}
		\meanBr{ v_ab_b\left(\dh_c\phi\right) }
		={}&
		- \frac{\lambda^2}{9} \meanBr{b^2} \bigg[
		\frac{1}{10} \delta_{bc} \dh_a h^v
		- \frac{7}{30} \delta_{ba} \dh_c h^v
		- \frac{1}{15} \delta_{ca}\dh_b h^v
		+ \frac{1}{6} \eps_{cab} \meanBr{\omega^2}
		\bigg] 
		+ \frac{\lambda^2}{108} \delta_{bc} h^c \dh_a\meanBr{ v^2 }
		\\& + \frac{\lambda^2}{216} \delta_{cb} h^v \dh_a\meanBr{ b^2 }
		+ \frac{5\lambda^2}{216} \delta_{ab} h^v \dh_c\meanBr{ b^2 }
		- \frac{5\lambda^2}{216} \delta_{ab}h^c \dh_c\meanBr{ v^2 }
		- \frac{\lambda^2}{216} \delta_{ac}h^c \dh_b\meanBr{ v^2 }
		- \frac{\lambda^2}{108} \delta_{ac} h^v \dh_b\meanBr{ b^2 }
		\\& - \frac{\lambda^2}{9} \meanBr{v^2} \bigg[
		\frac{7}{30} \delta_{ab} \dh_c h^c
		- \frac{1}{10} \delta_{ac} \dh_b h^c
		+ \frac{1}{15} \delta_{cb} \dh_a h^c
		+ \frac{1}{6} \eps_{cab} \meanBr{j^2}
		\bigg]
		+ \frac{2 \lambda^2}{54} \eps_{abc} h^v h^c
		.
	\end{split}
	\label{eq: vbdphi correlator without mean velocity}
\end{align}

For $a=c$, we can use \Eqsn{} \eqref{eq: vbdb correlator no derivatives or mean velocity} and \eqref{eq: vbddb correlator a=c} along with \Eq{eq: vbdphi correlator without mean velocity} to write
\begin{align}
	\begin{split}
		\meanBr{ v_cb_b\left(\dh_c\phi\right) }
		={}&
		\tau \eps_{klb} \frac{\lambda^2}{54} \left( \dh_k \mean{V}_l \right) \frac{1}{4\pi\rho} h^c h^c
		+ \tau \epsilon_{blm} \frac{\lambda^2}{54} \mean{V}_l \frac{1}{4\pi\rho} \meanBr{ b^2 } \dh_m \meanBr{j^2}
		- \tau \epsilon_{blj} \frac{\lambda^2}{216} \mean{V}_l \frac{1}{4\pi\rho} \meanBr{j^2} \dh_j\meanBr{ b^2 } 
		\\& + \tau \eps_{klb} \frac{\lambda^2}{54} \mean{V}_l \frac{1}{4\pi\rho}  h^c \dh_k h^c 
		- \frac{\tau\lambda^2}{54} \frac{1}{4\pi\rho} \mean{V}_b h^c \meanBr{j^2}
		+ \tau \epsilon_{jlb} \frac{\lambda^2}{18} \mean{V}_l \meanBr{j^2} \dh_j\meanBr{ v^2 }
		+ \tau \eps_{kbl} \frac{\lambda^2}{108} \mean{V}_l h^v \dh_k h^c 
		\\& + \tau \epsilon_{klb} \frac{\lambda^2}{54} \mean{V}_l \meanBr{ v^2 } \dh_k \meanBr{j^2}
		- \tau \epsilon_{bli} \frac{\lambda^2}{216} \mean{V}_l \meanBr{\omega^2} \dh_i\meanBr{ b^2 }
		+ \tau \eps_{bkl} \frac{\lambda^2}{27} \mean{V}_l h^c \dh_k h^v
		- \tau \frac{\lambda^2}{54} \mean{V}_b h^c \meanBr{\omega^2}
		\\& - \tau \epsilon_{blm} \frac{\lambda^2}{54} \mean{V}_l \meanBr{ b^2 } \dh_m \meanBr{\omega^2}
		+ \tau \frac{1}{4\pi\rho} \eps_{klb} \frac{\lambda^2}{54} \left( \dh_k \mean{V}_l \right) h^c h^c
		+ \tau \frac{1}{4\pi\rho} \eps_{blc} \frac{\lambda^2}{54} \left( \dh_c \mean{V}_l \right) \meanBr{j^2} \meanBr{ b^2 }
		\\& - \tau \eps_{blc} \frac{\lambda^2}{54} \left( \dh_c \mean{V}_l \right) \meanBr{\omega^2} \meanBr{ b^2 }
		- \tau \frac{1}{90} \eps_{klm} \frac{\lambda^2}{3} \left( \dh_c \mean{V}_l \right) 5\delta_{bk}\delta_{cm} \meanBr{\omega^2} \meanBr{ b^2 }
		+ \frac{\lambda^2}{27} \meanBr{b^2} \dh_b h^v
		\\& - \frac{\lambda^2}{36} h^c \dh_b\meanBr{ v^2 }
		.
	\end{split}
	\label{eq: vbdphi correlator a=c}
\end{align}

For $b=c$, we can use \Eqsn{} \eqref{eq: vbdb correlator no derivatives or mean velocity} and \eqref{eq: vbddb correlator b=c} along with \Eq{eq: vbdphi correlator without mean velocity} to write
\begin{align}
	\begin{split}
		\meanBr{ v_ab_c\left(\dh_c\phi\right) }
		={}&
		- \tau \eps_{kla} \frac{\lambda^2}{27} \left( \dh_k \mean{V}_l \right) \frac{1}{4\pi\rho} h^c h^c
		+ \tau \eps_{kla} \frac{\lambda^2}{18} \left( \dh_k \mean{V}_l \right)\frac{1}{4\pi\rho} \meanBr{j^2} \meanBr{ b^2 }
		- \tau \eps_{kla} \frac{\lambda^2}{18} \left( \dh_k \mean{V}_l \right) \meanBr{\omega^2} \meanBr{ b^2 }
		\\& - \tau \frac{\lambda^2}{54} \mean{V}_l \epsilon_{ila} \frac{1}{4\pi\rho} \meanBr{ b^2 } \dh_i \meanBr{j^2}
		- \tau \frac{\lambda^2}{27} \mean{V}_a \frac{1}{4\pi\rho} \meanBr{ b^2 } \int\d k\, 8 \pi k^4 N(k,\vec{R}) 
		\\& + \tau \frac{7\lambda^2}{180} \mean{V}_l \epsilon_{kla} \frac{1}{4\pi\rho} \meanBr{ b^2 }  \dh_k \meanBr{j^2}
		- \tau \frac{\lambda^2}{72} \mean{V}_l \epsilon_{alj} \frac{1}{4\pi\rho} \meanBr{j^2} \dh_j\meanBr{ b^2 }
		+ \tau \frac{\lambda^2}{54} \mean{V}_l \eps_{kal} \frac{1}{4\pi\rho} h^c \dh_k h^c
		\\& + \tau \frac{\lambda^2}{54} \mean{V}_a \frac{1}{4\pi\rho} h^c  \meanBr{j^2}
		- \tau \frac{\lambda^2}{24} \mean{V}_l \epsilon_{jla} \meanBr{\omega^2} \dh_j \meanBr{ b^2  }
		+ \tau \frac{\lambda^2}{27} \mean{V}_a \meanBr{ b^2 } \int\d k \, 8\pi k^4 F(k,\vec{R}) 
		\\& - \tau \frac{7\lambda^2}{180} \mean{V}_l \epsilon_{kla} \meanBr{ b^2 } \dh_k \meanBr{\omega^2}
		- \tau \frac{\lambda^2}{108} \mean{V}_l \epsilon_{kla} h^c \dh_k h^v
		+ \tau \frac{\lambda^2}{54} \mean{V}_a \meanBr{\omega^2} h^c
		+ \frac{\lambda^2}{36} h^v \dh_a\meanBr{ b^2 }
		\\& - \frac{\lambda^2}{27} \meanBr{v^2} \dh_a h^c
		.
	\end{split}
	\label{eq: vbdphi correlator b=c}
\end{align}

\subsubsection{\texorpdfstring{$\meanBr{ b_j\phi }$}{<bphi>}}
\label{section: appendix bphi correlator}
Using \Eq{eq: expression for phi}, we can write this correlator as
\begin{align}
	\begin{split}
		\meanBr{ b_j\phi } 
		={}& \eps_{klm}\int\d^3r\frac{1}{4\pi\left|\vec{r}\right|} 
		\mean{V}_l(\vec{x}+\vec{r}) \meanBr{ b_j(\vec{x}) \frac{\dh b_m(\vec{x}+\vec{r}) }{\dh r_k}  } 
		+ \eps_{klm}\int\d^3r\frac{1}{4\pi\left|\vec{r}\right|} 
		\frac{\dh \mean{V}_l(\vec{y}) }{\dh y_k} \meanBr{ b_j(\vec{x})b_m(\vec{x}+\vec{r})  } 
		\\& + \eps_{klm}\int\d^3r\frac{1}{4\pi\left|\vec{r}\right|}  
		\frac{\dh}{\dh r_k}\meanBr{ b_j(\vec{x})v_l(\vec{x}+\vec{r})b_m(\vec{x}+\vec{r}) } 
		.
	\end{split}
\end{align}
Using the Vishniac-Cho approximation, we write
\begin{align}
	\begin{split}
		\meanBr{ b_j\phi } 
		={}& \eps_{klm} \lambda^2 \mean{V}_l \meanBr{ b_j \dh_k b_m  } 
		+ \eps_{klm} \lambda^2 \left( \dh_k \mean{V}_l \right) \meanBr{ b_j b_m  } 
		+ \eps_{klm} \lambda^2 \dh_k \meanBr{   v_l b_j b_m   }
		- \eps_{klm} \lambda^2 \meanBr{  v_l b_m \left(\dh_k b_j\right)  } 
		.
	\end{split} \label{eq: bphi correlator intermediate step 1}
\end{align}
For the correlator $\meanBr{vbb}$, we can use \Eq{eq: vbb triple without pressure or large-scale derivatives}, while for the correlator $\meanBr{vb\dh b}$, we can use \Eq{eq: eps_klm <vbdb>}.
Using these along with \Eqsn{} \eqref{eq: bibj correlator} and \eqref{eq: bidkbj correlator} and keeping only up to one large-scale derivative, the above can be written as
\begin{align}
	\begin{split}
		\meanBr{ b_j\phi } 
		={}& 
		\lambda^2 \frac{1}{12} \eps_{jkl} \mean{V}_l \dh_k\meanBr{ b^2 } 
		- \overbrace{ \lambda^2 \frac{1}{3} h^c \mean{V}_j }^{\mathclap{\text{part of the advective flux}}} {}
		+ \lambda^2\frac{1}{3} \meanBr{ b^2 } \eps_{jkl}  \left( \dh_k \mean{V}_l \right)
		+ \tau^2\lambda^2 \frac{1}{18} \frac{1}{4\pi\rho}  h^c  h^c \epsilon_{jdk} \left(\dh_d\mean{V}_k \right)
		\\& - \tau\lambda^2 \frac{1}{4\pi\rho} \frac{1}{6} h^c \dh_j\meanBr{ b^2 } 
		+ \tau\lambda^2 \frac{1}{4\pi\rho} \frac{1}{9} \meanBr{ b^2 } \dh_j h^c
		- \tau\lambda^2 \frac{1}{12} h^v \dh_j\meanBr{ b^2 }
		- \tau\lambda^2 \frac{5}{36} h^c \dh_j\meanBr{ v^2 } 
		+ \tau\lambda^2 \frac{1}{9} \meanBr{ v^2 } \dh_j h^c
		\,.
	\end{split} \label{eq: bphi correlator contribution}
\end{align}
We note that if one is willing to make the identification $\lambda^2 h^c = h^b$, the second term above is just a part of the already-known advective flux (we show this in appendix \ref{appendix: advective flux}).

\subsubsection{\texorpdfstring{$\meanBr{ \left( \dh_a\phi\right) \dh_d \left( v_b b_c\right) }$}{<DphiD(vb)>}}
Here, we do not need to keep terms containing the mean velocity, but we need to keep up to one large-scale derivative.
Using \Eq{eq: expression for phi} and following a procedure similar to that used to obtain \Eq{eq: dhpba correlator before VC}, we write
\begin{align}
	\begin{split}
		\meanBr{ \left( \dh_a\phi\right) \dh_d \left( v_b b_c\right) }
		={}&
		\eps_{klm} \int\frac{\d^3r}{4\pi} \left( - \frac{\delta_{ka}}{r^3} + \frac{3r_k r_a}{r^5} \right) \meanBr{ \frac{\dh v_b(\vec{x})}{\dh x_d} v_l(\vec{x} - \vec{r}) } \meanBr{ b_c(\vec{x}) b_m(\vec{x}-\vec{r})  }
		\\& + \eps_{klm} \int\frac{\d^3r}{4\pi} \left( - \frac{\delta_{ka}}{r^3} + \frac{3r_k r_a}{r^5} \right) \meanBr{ v_b(\vec{x}) v_l(\vec{x} - \vec{r}) } \meanBr{ \frac{\dh b_c(\vec{x}) }{\dh x_d}  b_m(\vec{x}-\vec{r})  }
		.
	\end{split}
\end{align}
Using the Vishniac-Cho approximation, we write, for $a=b$ and $a=c$ respectively,
\begin{align}
	\begin{split}
		\meanBr{ \left( \dh_a\phi\right) \dh_d \left( v_a b_c\right) }
		\approx{}&
		- \lambda^2 \frac{1}{36} \eps_{cdk} h^c \dh_k h^v
		+ \lambda^2 \frac{1}{360} \eps_{dac} \meanBr{\omega^2} \dh_a\meanBr{ b^2 }
		+ \lambda^2 \frac{1}{90} \delta_{cd} h^c \meanBr{\omega^2}
		\\& - \eps_{ldc} \lambda^2 \frac{7}{1080} \meanBr{j^2}\dh_l\meanBr{ v^2 } 
		- \lambda^2 \frac{1}{540} \eps_{cdg} h^v \dh_g h^c
        - \lambda^2 \frac{1}{270} \delta_{cd} h^v \meanBr{j^2}
		\\& + \lambda^2 \frac{1}{135} \delta_{cd} \meanBr{b^2} \int\d k \, 8\pi k^4 F(k,\vec{R})
		+ \lambda^2 \frac{13}{1350} \eps_{dac} \meanBr{b^2} \dh_a \meanBr{\omega^2},
		\label{eq: (dphi)d(vb) correlator a=b}
	\end{split}
\end{align}
\begin{align}
	\begin{split}
		\meanBr{ \left( \dh_a\phi\right) \dh_d \left( v_b b_a\right) }
		\approx{}&
		\eps_{dbm} \lambda^2 \frac{7}{1080} \meanBr{\omega^2} \dh_m\meanBr{ b^2 }
		+ \lambda^2 \frac{1}{36} \eps_{lbd} h^v \dh_l h^c
		+ \lambda^2 \frac{1}{360} \eps_{dba} \meanBr{j^2} \dh_a\meanBr{ v^2 }
		\\& - \lambda^2 \frac{1}{90} \delta_{bd} h^v \meanBr{j^2}
		- \lambda^2 \frac{1}{135} \delta_{bd} \meanBr{v^2} \int\d k\, 8 \pi k^4 N(k,\vec{R})
		\\& + \eps_{dbm} \lambda^2 \frac{22}{2025} \meanBr{v^2} \dh_m \meanBr{j^2}
		+ \lambda^2 \frac{1}{540} \eps_{bde} h^c \dh_e h^v
		+ \lambda^2 \frac{1}{270} \delta_{bd} h^c \meanBr{\omega^2}.
		\label{eq: (dphi)d(vb) correlator a=c}
	\end{split}
\end{align}

\subsection{More triple correlators}
\label{appendix: calc triple correlator with derivative inside}

\subsubsection{\texorpdfstring{$\meanBr{v_ab_bb_c}$}{<vbb>}}
Using \Eqsn{} \eqref{eq: bibj correlator}, \eqref{eq: bidjdkbl correlator}, \eqref{eq: vivj correlator}, and \eqref{eq: vidkvj correlator}, dropping the nonlocal pressure term, and keeping no large-scale derivatives, we write \Eq{eq: vbb triple} as
\begin{align}
	\begin{split}
		\meanBr{ v_ab_bb_c } = \bigO(\dh)
		\,,
	\end{split} \label{eq: vbb triple without pressure or large-scale derivatives}
\end{align}
i.e.\@ all the local terms involve large-scale derivatives.

\subsubsection{\texorpdfstring{$\meanBr{ v_l\left(\dh_k b_j\right)b_m }$}{v(Db)b}}
\label{appendix: vdbb correlator}
We need the combination $\eps_{klm} \meanBr{v_l b_m \dh_k b_j}$ (which appears in \Eq{eq: bphi correlator intermediate step 1}).
We do not need to keep any nonlocal terms, and we need to keep up to one large-scale derivative.
We can throw away $\bigO(\tau^3)$ terms.
Following the same procedure as in appendix \ref{appendix: triple correlator evaluation}, we obtain
\begin{align}
	\begin{split}
		\eps_{klm} \meanBr{ v_l\left(\dh_k b_j\right)b_m }
		={}&
		- \tau^2 \frac{1}{18} \epsilon_{jdk} \left(\dh_d\mean{V}_k \right) \frac{1}{4\pi\rho}  h^c  h^c
		+ \tau \frac{1}{4\pi\rho} \frac{1}{6} h^c \dh_j\meanBr{ b^2 } 
		- \tau \frac{1}{4\pi\rho} \frac{1}{9} \meanBr{ b^2 } \dh_j h^c
		\\& + \tau \frac{1}{12} h^v \dh_j\meanBr{ b^2 }
		+ \tau \frac{5}{36} h^c \dh_j\meanBr{ v^2 } 
		- \tau \frac{1}{9} \meanBr{ v^2 } \dh_j h^c
		+ \bigO(\tau^3)
		\,.
	\end{split}
	\label{eq: eps_klm <vbdb>}
\end{align}

A similar correlator also appears in \Eq{eq: vbdphi correlator intermediate 1}, but this time, we can discard terms involving the mean velocity, terms with large-scale derivatives, and nonlocal terms.
Following the same procedure as in appendix \ref{appendix: triple correlator evaluation}, we obtain
\begin{align}
	\begin{split}
		\meanBr{ v_lb_m\left(\dh_k b_j\right) }
		={}&
		\tau \frac{1}{4\pi\rho} \frac{1}{36} h^c h^c \left( \delta_{kl}\delta_{jm} - \delta_{km}\delta_{jl} \right)
		+ \tau \frac{1}{4\pi\rho} \frac{1}{90} \left( 4\delta_{mk}\delta_{lj} - \delta_{ml}\delta_{kj} - \delta_{mj}\delta_{kl} \right) \meanBr{j^2} \meanBr{ b^2 }
		\\& - \tau \frac{1}{90} \left( 4\delta_{mk}\delta_{lj} - \delta_{ml}\delta_{kj} - \delta_{mj}\delta_{kl} \right) \meanBr{\omega^2} \meanBr{ b^2 }
		+ \bigO(\tau^2)
		\,.
		\end{split}
		\label{eq: vbdb correlator no derivatives or mean velocity}
\end{align}

\subsubsection{\texorpdfstring{$\meanBr{ b_b a_c \dh_e \dh_a v_d }$}{<baDDv>}}
This correlator appears multiplied with the mean velocity, and we don't need to keep any large-scale derivatives in it.
This correlator also appears multiplied with the square of the correlation length, so we will not keep any nonlocal terms (involving the pressure and the scalar potential).
Following the same procedure as in appendix \ref{appendix: triple correlator evaluation}, we write
\begin{align}
	\begin{split}
		\meanBr{ \left(\dh_e\dh_iv_d\right)b_ba_i }
		={}&
		\frac{\tau}{4\pi\rho} \meanBr{  \left(\dh_e\dh_ib_j\right) b_b } \epsilon_{ijd} \frac{\meanBr{b^2}}{6}
		+ \frac{\tau}{4\pi\rho} \epsilon_{bij} \frac{h^c}{6} \meanBr{ \left(\dh_e\dh_jb_d \right) a_i }
		+ \frac{\tau}{4\pi\rho} \frac{1}{3} h^b \meanBr{ \left(\dh_e \nabla^2 b_d\right)  b_b }
		\\& - \tau \meanBr{ \left(\dh_e\nabla^2 v_d\right) v_b } \frac{1}{3} h^b
		+ \tau \meanBr{ \left(\dh_e\dh_iv_d\right) v_j } \epsilon_{ijb} \frac{\meanBr{b^2}}{6}
		+ \bigO(\tau^2)
	\end{split} \label{eq: baddv triple correlator a equal c}
\end{align}
and
\begin{align}
	\begin{split}
		\meanBr{ \left(\dh_e\dh_av_d\right)b_ia_i }
		={}&
		\frac{\tau}{4\pi\rho} \meanBr{  \left(\dh_e\dh_ab_j\right) b_i }  \epsilon_{ijd} \frac{\meanBr{b^2}}{6}
		+ \frac{\tau}{4\pi\rho} \meanBr{ \left(\dh_e\dh_ab_j\right) a_i } \epsilon_{ijd} \frac{h^c}{6}
		+ \frac{\tau}{4\pi\rho} \epsilon_{iaj} \frac{h^c}{6} \meanBr{ \left(\dh_e\dh_jb_d \right) a_i }
		\\& + \frac{\tau}{4\pi\rho} \epsilon_{iaj} \frac{\meanBr{b^2}}{6} \meanBr{ \left(\dh_e\dh_jb_d \right)  b_i }
		+ \frac{\tau}{4\pi\rho} \epsilon_{iej} \frac{h^c}{6} \meanBr{ \left(\dh_a\dh_jb_d\right) a_i }
		+ \frac{\tau}{4\pi\rho} \epsilon_{iej} \frac{\meanBr{b^2}}{6} \meanBr{ \left(\dh_a\dh_jb_d\right) b_i }
		+ \bigO(\tau^2)
		\,.
	\end{split}  \label{eq: baddv triple correlator b equal c}
\end{align}

\subsubsection{\texorpdfstring{$\meanBr{ v_a b_b \dh_k \dh_c b_m }$}{<vbDDb>}}
This correlator appears in \Eq{eq: vbdphi correlator intermediate 1}, multiplied by $\epsilon_{klm}$. 
We need to keep up to one large-scale derivative, but we can discard terms involving the mean velocity.
It is also multiplied by the square of the correlation length, so we will not keep any nonlocal terms.
Following the same procedure as in appendix \ref{appendix: triple correlator evaluation}, we write
\begin{align}
	\begin{split}
		\epsilon_{klm} \meanBr{ v_i b_b \left( \dh_k \dh_i b_m\right) }
		={}&
		\tau \epsilon_{blm} \frac{1}{4\pi\rho} \frac{1}{18} \meanBr{ b^2 } \dh_m \meanBr{j^2}
		- \tau \epsilon_{blj} \frac{1}{4\pi\rho} \frac{1}{72} \meanBr{j^2} \dh_j\meanBr{ b^2 } 
		+ \tau \eps_{bkl} \frac{1}{4\pi\rho}  \frac{1}{18} h^c \dh_k h^c 
		\\& - \tau \delta_{lb} \frac{1}{4\pi\rho}  \frac{1}{18} h^c \meanBr{j^2}
		+ \tau \frac{1}{6} \epsilon_{jlb} \meanBr{j^2} \dh_j\meanBr{ v^2 }
		+ \tau \eps_{kbl} \frac{1}{36} h^v \dh_k h^c 
		+ \tau \epsilon_{klb} \frac{1}{18} \meanBr{ v^2 } \dh_k \meanBr{j^2}
		\\& - \tau \epsilon_{bli} \frac{1}{72} \meanBr{\omega^2} \dh_i\meanBr{ b^2 }
		+ \tau \frac{1}{9} \eps_{bkl} h^c \dh_k h^v
		- \tau \delta_{lb} \frac{1}{18} h^c \meanBr{\omega^2}
		- \tau \epsilon_{blm} \frac{1}{18} \meanBr{ b^2 } \dh_m \meanBr{\omega^2}
	\end{split}
	\label{eq: vbddb correlator a=c}
\end{align}
\begin{align}
	\begin{split}
		\epsilon_{klm} \meanBr{ v_a b_i \left( \dh_k \dh_i b_m\right) }
		={}&
		- \tau \epsilon_{ila} \frac{1}{4\pi\rho} \frac{1}{18} \meanBr{ b^2 } \dh_i \meanBr{j^2}
		- \tau \delta_{la} \frac{1}{4\pi\rho} \frac{1}{9} \meanBr{ b^2 } \int\d k\, 8 \pi k^4 N(k,\vec{R}) 
		\\& + \tau \epsilon_{kla} \frac{1}{4\pi\rho} \frac{7}{60}\meanBr{ b^2 }  \dh_k \meanBr{j^2}
		- \tau \epsilon_{alj} \frac{1}{4\pi\rho} \frac{1}{24} \meanBr{j^2} \dh_j\meanBr{ b^2 }
		+ \tau \eps_{kal} \frac{1}{4\pi\rho} \frac{1}{18} h^c \dh_k h^c
		\\& + \tau \delta_{al} \frac{1}{4\pi\rho} \frac{1}{18} h^c  \meanBr{j^2}
		- \tau \epsilon_{jla} \frac{1}{8} \meanBr{\omega^2} \dh_j \meanBr{ b^2  }
		+ \tau \delta_{la} \frac{1}{9} \meanBr{ b^2 } \int\d k \, 8\pi k^4 F(k,\vec{R}) 
		\\& - \tau \epsilon_{kla} \frac{7}{60} \meanBr{ b^2 } \dh_k \meanBr{\omega^2}
		- \tau \epsilon_{kla} \frac{1}{36} h^c \dh_k h^v
		+ \tau \delta_{al} \frac{1}{18} \meanBr{\omega^2} h^c
	\end{split}
	\label{eq: vbddb correlator b=c}
\end{align}

\section{The advective flux}
\label{appendix: advective flux}
Let us try to evaluate the contribution of the $-\meanvec{V}\times\vec{b}$ term of $\vec{e}$ (\Eq{eq: e definition}).
Writing $e_k = -\eps_{klm}\mean{V}_lb_m$, we can write the last term on the RHS of \Eq{eq: hb evolution in terms of e} as
\begin{equation}
	- \dh_j\meanBr{ \eps_{jki}e_ka_i }\bigg|_{advective} 
	= \dh_j\Big[\mean{V}_i\meanBr{ b_ja_i }\Big] - \dh_j\Big[\mean{V}_j\meanBr{ b_ia_i }\Big]
	.
\end{equation}
Using \Eq{eq: aibj correlator}, we can write the above as
\begin{equation}
	- \dh_j\meanBr{ \eps_{jki}e_ka_i }\bigg|_{advective} 
	= -\frac{2}{3}\dh_j\Big[\mean{V}_jh^b\Big] + \frac{1}{12} \epsilon_{ijm} \left( \dh_j \mean{V}_i \right)  \dh_m \meanBr{a^2}
	.
	\label{eq: advective flux extra term}
\end{equation}
Recall that in appendix \ref{section: appendix bphi correlator}, we had obtained a similar advective contribution from the $\meanBr{ b_i\dh_i\varphi }$ term (\Eq{eq: bphi correlator contribution}).
Adding both these terms (and identifying $\lambda^2 h^c = h^b$), we recover an advective flux given by $V_j h^b$, in agreement with the gauge-invariant calculation \citep{Kandu2006helicityFlux}.

\section{Full expression without further approximations}
\label{appendix: dhbdt nonlocal with hv and lam^2}
\begin{align}
	\begin{split}
		-\dh_j(F^T_j)
		={}& 
		\dh_j\Bigg[
			\frac{\tau}{9} \left( \meanBr{v^2} + \frac{\meanBr{b^2}}{4\pi\rho} \right) \dh_j h^b
			+ \tau \frac{1}{18} \left( h^v - \frac{h^c}{4\pi\rho} \right) \dh_j\meanBr{a^2}
			+ \frac{\tau}{9} h^b \dh_j \left( \meanBr{ v^2 } - \frac{\meanBr{ b^2 }}{4\pi\rho} \right)
			\\&\continuedTerm - \tau\lambda^2 \frac{5}{54} \frac{1}{4\pi\rho} h^c \dh_j\meanBr{ b^2 } 
			+ \tau\lambda^2 \frac{1}{4\pi\rho} \frac{5}{27} \meanBr{ b^2 } \dh_j h^c
			- \tau\lambda^2 \frac{1}{9} h^v \dh_j\meanBr{ b^2 }
			- \tau\lambda^2 \frac{1}{6} h^c \dh_j\meanBr{ v^2 } 
			\\&\continuedTerm + \tau\lambda^2 \frac{4}{27} \meanBr{ v^2 } \dh_j h^c
			- \tau \frac{\lambda^2}{27} \frac{1}{4\pi\rho} \meanBr{j^2} \dh_j h^b 
			+ \frac{\tau\lambda^2}{27} \meanBr{b^2} \dh_j h^v
			+ \frac{\lambda^2 }{4} \meanBr{ b^2 } \eps_{jkl} \left( \dh_k \mean{V}_l \right)
			\\&\continuedTerm - \frac{\tau^2 \lambda^2}{27} \left( \dh_d \mean{V}_e \right) \frac{1}{4\pi\rho} \eps_{jde} h^b \int\d k \, 8\pi k^4 N(k,\vec{R})
			+ \frac{\tau^2 \lambda^2}{27} \left( \dh_d \mean{V}_e \right) \eps_{jde} h^b \int\d k \, 8\pi k^4 F(k,\vec{R}) 
			\\&\continuedTerm + \frac{4 \tau^2\lambda^2}{27} \epsilon_{jde} \left( \dh_d \mean{V}_e \right) \frac{1}{4\pi\rho} h^c h^c
			- \tau^2 \frac{\lambda^2}{36} \epsilon_{jlk} \mean{V}_l \frac{1}{4\pi\rho} \meanBr{j^2} \dh_k\meanBr{ b^2 } 
			- \tau^2 \frac{4\lambda^2}{135} \eps_{jlk} \mean{V}_l \frac{1}{4\pi\rho}  h^c \dh_k h^c
			\\&\continuedTerm + \tau^2 \frac{7 \lambda^2}{108} \epsilon_{jel} \mean{V}_l \meanBr{j^2} \dh_e\meanBr{ v^2 }
			- \tau^2 \frac{7 \lambda^2}{180} \eps_{jkl} \mean{V}_l h^v \dh_k h^c 
			+ \tau^2 \frac{119 \lambda^2}{4050} \epsilon_{jkl} \mean{V}_l \meanBr{ v^2 } \dh_k \meanBr{j^2}
			\\&\continuedTerm - \tau^2 \frac{\lambda^2}{18} \epsilon_{jli} \mean{V}_l \meanBr{\omega^2} \dh_i\meanBr{ b^2 }
			+ \tau^2 \frac{\lambda^2}{60} \eps_{jkl} \mean{V}_l h^c \dh_k h^v
			- \tau^2 \frac{181 \lambda^2}{2700} \epsilon_{jlm} \mean{V}_l \meanBr{ b^2 } \dh_m \meanBr{\omega^2}
			\\&\continuedTerm - \tau^2 \frac{\lambda^2}{18} \frac{1}{4\pi\rho} \eps_{ejd} \left( \dh_d \mean{V}_e \right) \meanBr{j^2} \meanBr{ b^2 }
			- \tau^2 \frac{2\lambda^2}{27} \eps_{edj} \left( \dh_d \mean{V}_e \right) \meanBr{\omega^2} \meanBr{ b^2 }
			\\&\continuedTerm - \tau^2 \frac{7\lambda^2}{180} \mean{V}_l \epsilon_{klj} \frac{1}{4\pi\rho} \meanBr{ b^2 }  \dh_k \meanBr{j^2}
			- \mean{V}_l \tau^2 \eps_{jli} \frac{1}{18} \frac{1}{4\pi\rho} \meanBr{b^2} \dh_i\meanBr{ b^2 }
			+ \mean{V}_l \tau^2 \eps_{jli} \frac{1}{6} \meanBr{b^2} \dh_i\meanBr{ v^2 } 
			\\&\continuedTerm + \mean{V}_l \tau^2 \eps_{jli} \frac{1}{9} \meanBr{v^2} \dh_i\meanBr{b^2}
			+ \mean{V}_l \tau^2 \eps_{jli} \frac{1}{18} \frac{h^c}{4\pi\rho} \dh_i h^b
			- \mean{V}_l \tau^2 \eps_{jli} \frac{1}{18} h^v \dh_i h^b 
		\Bigg]
		\\& - \frac{1}{3} \dh_j\Big[ \lambda^2 h^c \mean{V}_j \Big]
		-\frac{2}{3}\dh_j\Big[ h^b \mean{V}_j\Big] + \frac{1}{12} \epsilon_{ijm} \left( \dh_j \mean{V}_i \right)  \dh_m \meanBr{a^2}
	\end{split}
\end{align}

\bibliography{helicity-fluxes-from-triple}{}
\bibliographystyle{aasjournal}


\end{document}